\documentclass[fleqn,usenatbib]{mnras}

\usepackage{newtxtext,newtxmath}
\usepackage[T1]{fontenc}
\usepackage{graphicx}
\usepackage{amsmath}
\usepackage{siunitx}
\usepackage{bm}

\newcommand{\logNHI}{\ensuremath{\log N(\mbox{\ion{H}{i}})}}
\newcommand{\logNHIunit}{\ensuremath{\log [N(\mbox{\ion{H}{i}})/\rm{cm}^{-2}]}}
\newcommand{\kms}{\ensuremath{\text{km s}^{-1}}}

\DeclareUnicodeCharacter{0308}{}

\title[The physical origins of absorbers]{The physical origins of gas in the circumgalactic medium using observationally-motivated TNG50 mocks}

\author[Weng et al.]{Simon Weng,$^{1,2,3,4}$\thanks{E-mail: simonw358@gmail.com}
{C\'eline P\'eroux},$^{1,5}$
Rahul Ramesh,$^{6}$
Dylan Nelson,$^{6}$ 
Elaine M. Sadler,$^{2, 3, 4}$
\newauthor
Martin Zwaan,$^{1}$ 
Victoria Bollo,$^{1}$
Benedetta Casavecchia,$^{7}$
\\
$^{1}$ European Southern Observatory, Karl-Schwarzschildstrasse 2, D-85748 Garching bei M{\"u}nchen, Germany\\
$^2$ Sydney Institute for Astronomy, School of Physics A28, University of Sydney, NSW 2006, Australia\\
$^3$ ARC Centre of Excellence for All Sky Astrophysics in 3 Dimensions (ASTRO 3D)\\ 
$^4$ ATNF, CSIRO Space and Astronomy,  PO Box 76, Epping, NSW 1710, Australia \\
$^5$ Aix Marseille Universit\'e, CNRS, LAM (Laboratoire d'Astrophysique de Marseille) UMR 7326, 13388, Marseille, France \\
$^6$ Universit\"{a}t Heidelberg, Zentrum f\"{u}r Astronomie, Institut f\"{u}r theoretische Astrophysik, Albert-Ueberle-Str. 2, 69120 Heidelberg, Germany\\
$^{7}$ Max-Planck-Institut f\"{u}r Astrophysik, Karl-Schwarzschild-Strasse 1, D-85748 Garching b. M\"{u}nchen, Germany
}

\date{Accepted XXX. Received YYY; in original form ZZZ}

\pubyear{2023}

\begin{document}
\label{firstpage}
\pagerange{\pageref{firstpage}--\pageref{lastpage}}
\maketitle

\begin{abstract}
Absorbers in the spectrum of background objects probe the circumgalactic medium (CGM) surrounding galaxies, but its physical properties remain unconstrained. 
We use the cosmological hydrodynamical simulation TNG50 to statistically trace the origins of \ion{H}{i} Ly-$\alpha$ absorbers around galaxies at $z = 0.5$ with stellar masses ranging from 10$^8$ to 10$^{11}$ M$_\odot$.  
We emulate observational CGM studies by considering all gas within a line of sight velocity range of $\pm 500$ \kms\ from the central, to quantitatively assess the impact of other galaxy haloes and overdense gas in the IGM that intersect sightlines. 
We find that 75 per cent of \ion{H}{i} absorbers with column densities $\logNHIunit > 16.0$ trace the central galaxy within $\pm150$ (80) \kms\ of $M_* = 10^{10} (10^8)$ M$_\odot$ central galaxies. 
The impact of satellites to the total absorber fraction is most significant at impact parameters $0.5 R_{\rm vir} < b < R_{\rm vir}$, and satellites with masses below typical detection limits ($M_* < 10^8$ M$_\odot$) account for 10 (40) per cent of absorbers that intersect any satellite bound to $10^{10}$ and $10^{11}$ $(10^9)$ M$_\odot$ centrals. 
After confirming outflows are more dominant along the minor axis, we additionally show that at least 20 per cent of absorbers exhibit no significant radial movement, indicating that absorbers can also trace quasi-static gas. 
Our work shows that determining the stellar mass of galaxies at $z_{\rm abs}$ is essential to constrain the physical origin of the gas traced in absorption, which in turn is key to characterising the kinematics and distribution of gas and metals in the CGM.   
\end{abstract}



\section{Introduction}
The discovery of extragalactic absorption lines \citep[e.g.][]{Lynds1966, Burbidge1966, Stockton1966} shortly followed the discovery of the first quasar/quasi-stellar object (QSO) 3C 273 \citep{Schmidt1963}. 
At the time, it was suggested that the absorption arose from discrete clouds of gas found in galaxies located along the line of sight (LOS) towards the background source \citep{Bahcall1965}. 
Almost six decades on, the study of absorbers has evolved significantly but fundamental questions remain unanswered. 
One such question concerns the origin of absorbers; while absorption lines are commonly associated with galaxies near the line-of-sight \citep{Bahcall1969, Bergeron1986}, what is their spatial distribution, kinematic behaviour and origin of the gas being traced in and around galaxies? 

The distribution of gas in and around galaxies remains in a constant state of flux. 
Like a galactic bank account, reservoirs of gas are depleted by constant withdrawals in the form of star formation. 
More sudden ejections of gas in the form of stellar-wind-driven and AGN-driven outflows further reduce the amount of gas available for star formation but also enrich the surrounding galaxy halo \citep{Veilleux2005}. 
Deposits in the form of accreting gas from dark matter filaments \citep{Keres2005, Keres2009, Nelson2013}, condensed material from galactic fountains \citep{Fraternali2008, Marinacci2010, Fraternali2017} and `clumpy' accretion in the form of satellites \citep{Mihos1994, Hernquist1995, Ramon2020} replenish the gas supply. 
These processes take place simultaneously in a delicate balance and affect the stellar properties of galaxies. 
The study of the circumgalactic medium (CGM), the region where inflows and outflows leave their signatures, is of paramount importance to understanding how galaxies form and evolve.

Intervening quasar absorbers have long been found at the same redshift of galaxies and intersecting gaseous haloes around galaxies \citep{Bahcall1969, Bergeron1986}. 
The circumgalactic medium that extends from the interstellar medium (ISM) to the intergalactic medium (IGM) can be traced by absorption lines towards background QSOs. 
Some surveys study the CGM by preselecting galaxies (using a property such as mass or luminosity) that are located near quasar sightlines and then searching for and analysing absorption lines in the QSO spectrum \citep[e.g.][]{Chen2010, Tumlinson2011, Werk2016, Heckman2017, Johnson2017, Berg2018, Muzahid2018, Pointon2019}. 
Alternatively, one can identify absorption lines towards a distant background source, typically of a certain species such as \ion{H}{i} or \ion{Mg}{ii}, and then search for galaxies at the absorber redshift \citep[e.g.][]{Steidel1992, Schroetter2016, Hamanowicz2020, Lofthouse2020, Muzahid2020, Nielsen2020, Peroux2022, Banerjee2023, Berg2023}. 
Finally, surveys can target fields with a UV-bright QSO without a pre-selection on absorption lines or galaxies \citep[e.g.][]{ Prochaska2011, Prochaska2019, Burchett2019, Chen2020, Muzahid2021}. 
Despite the differing sample selections, surveys of absorption lines and their associated galaxies have benefited from the advent of integral field spectroscopy (IFS). 
The ability to simultaneously obtain imaging and spectroscopy has led to a proliferation of galaxy--absorber pairs \citep[e.g. $>100$ for \ion{Mg}{ii};][]{Schroetter2019, Dutta2020}. 

It is then inevitable that absorption lines in QSO spectra will intersect gas associated with processes such as inflows and outflows or structures like satellites and filaments \citep{Lehner2013, Lehner2019, Lehner2022, Berg2023}. 
However, disentangling the relationship between absorber and galaxy and understanding the origin of the absorber becomes difficult for many reasons. 
First and foremost, we are unable to directly measure the distance between absorber and galaxy in the direction along the line of sight. 
The only indirect measure of distance is the line of sight velocity difference between the galaxy and absorber ($\Delta v_{\rm LOS}$). 
However, this is determined by a combination of the Hubble flow and peculiar velocity of the gas cloud. 
Several studies using simulations show that selecting absorbers in velocity space can select for gas beyond several times the virial radius \citep{Rahmati2015, Ho2020, Ho2021}.
Identifying the origin of an absorber becomes further complicated in studies where galaxy overdensities are found associated with absorbers \citep[e.g.][]{Gauthier2013, Bielby2017, Manuwal2019, Hamanowicz2020}. 
Typically, the galaxy found at closest impact parameter to the QSO sightline is selected as the absorber host, but other methods have also been used \citep{Berg2023}. 
Finally, the inherent sensitivity limit in observations means that a population of low-mass and/or quiescent galaxies will remain undetected, particularly at cosmological redshifts. 
We can infer the existence of these galaxies from studies that do not detect any object near high column density absorbers that are expected to arise from the galaxy disk \citep[e.g.][]{Fumagalli2015, Weng2023a}. 
The combination of these factors means that assigning an origin to absorbers is challenging but few works have attempted to quantify these effects and how they vary with properties such as species and column density. 

Intervening absorption lines are expected to probe gas flows into and out of galaxies. 
However, distinguishing between outflows and inflows is challenging because of the line of sight velocity degeneracy between the two gas flows. 
Gas infalling onto a galaxy from behind and gas outflowing towards the observer will both be blueshifted with respect to the systemic redshift of the galaxy. 
Likewise, redshifted absorbers can originate from inflows in front of the galaxy or outflows ejecting gas away from the observer. 
This degeneracy is not present in down-the-barrel studies where inflows (outflows) are identified by redshifted (blueshifted) absorption against the background stellar continuum \citep[e.g.][]{Martin2005, Veilleux2005, Rubin2014, Heckman2017}. 
To overcome this uncertainty in transverse absorption-line studies, one must rely on other assumptions. 

One possible way to distinguish outflowing absorbers is to measure the azimuthal angle ($\Phi$) between the absorber and a galaxy's major axis. 
In both observations and simulations of gas outflows that originate in the galaxy centre, galactic winds commonly form an expanding biconical shape perpendicular to the galaxy disk as this is the path of least resistance \citep[e.g.][]{Bordoloi2011, Lan2014, Nelson2019a}. 
Hence, one could assume that outflowing absorbers can be identiifed by their alignment with the minor axis and inflowing absorbers with the major axis as the accreting gas co-rotates with the galaxy \citep{Rahmani2018b, Schroetter2019, Zabl2021}. 
Perhaps further evidence of this major-minor axis dichotomy can be found in the bimodal distribution of azimuthal angles for absorbers relative to their galaxy hosts \citep{Bouche2012, Kacprzak2012, Bordoloi2014, Schroetter2019}. 
It should follow then that metal-enriched gas will be preferentially found near the minor axis. 
Indeed, various simulations predict an azimuthal angle dependence in metallicity profiles \citep{Peroux2020, vandevoort2021}, as well as density, temperature \citep{Truong2021, Yang2023} and magnetic field strengths \citep{Ramesh2023c}, in line with some observational studies \citep{Cameron2021, Zhang2022, Heesen2023}, although other observations find little to no evidence for such angular anisotropies \citep{Peroux2016, Kacprzak2019, Pointon2019, Huang2021, Wendt2021}. 

We also expect outflowing gas to be metal-enriched as heavy elements from the ISM are ejected into the CGM and IGM via feedback processes \citep{Oppenheimer2006, Muratov2017, Nelson2019a, Mitchell2022}. 
Studies of high-velocity clouds (HVCs) in the Milky Way use a combination of kinematics and metallicity to distinguish between inflows and outflows, and also calculate the mass rates of both processes \citep[e.g.][]{Wakker1999, Fox2004, Fox2016, Fox2019, Ramesh2023b}. 
Beyond the local Universe, some studies of \ion{H}{i} absorbers find a multimodal gas-phase metallicity distribution \citep{Lehner2013, Lehner2022}. 
The separate populations have been proposed to trace outflows (metal-rich), inflows (metal-poor) and gas in the IGM (pristine) \citep{Berg2023}. 
However, simulations do not find such large metallicity contrasts between inflows and outflows, particularly at lower redshift where galactic winds are more efficiently recycled \citep{Hafen2017}. 

In this work, we study the origins of absorbers using the cosmological magnetohydrodynamical simulation TNG50. 
We explore how the line of sight velocity difference between galaxy and absorber compares with the physical distance.  
Then, we quantify the contributions of satellite galaxies and gas in the intergalactic medium to the fraction of absorbers with varying column densities. 
We also test the fidelity of azimuthal angle and metallicity assumptions when identifying gas flows in the CGM. 
Finally, we compare our results to studies of Ly-$\alpha$ absorbers at redshift $z \lesssim 1$ such as the MUSE-ALMA Haloes survey \citep{Peroux2022} and the COS CGM Compendium \citep{Lehner2018}. 
{We adopt a cosmology consistent with \citet{Planck2016} and halo masses, circular velocities and radii are defined at 200 times the critical density of the Universe (e.g. $R_{\rm vir} = R_{\rm 200c}$), consistent with TNG50.} 


\section{Methods}
\label{sec:2methods}

\subsection{The TNG50 Simulation}

We present results from TNG50-1 \citep{Nelson2019a, Pillepich2019}, the highest resolution version in the IllustrisTNG (henceforth, TNG) suite of cosmological magneto-hydrodynamical simulations \citep{Marinacci2018, Naiman2018, Nelson2018, Pillepich2018b, Springel2018}. Building on the original Illustris simulation \citep{Genel2014, Vogelsberger2014a, Vogelsberger2014b, Sijacki2015}, IllustrisTNG incorporates magnetic fields \citep{Pakmor2014} and amends the original Illustris galaxy formation model \citep{Vogelsberger2013, Torrey2014} with updated feedback processes \citep{Weinberger2017, Pillepich2018a}. 

With a box size of $\sim$50 comoving Mpc (cMpc) and 2160$^3$ resolution elements, the TNG50 simulation has a baryonic (dark matter) mass resolution of $8.5 \times 10^4$ M$_\odot$ ($4.5 \times 10^5$ M$_\odot$). The relatively large volume combined with the high spatial resolution are optimal for the study of structures in the circumgalactic medium of galaxies. 

In the TNG simulations, galaxy stellar mass growth is regulated by supernovae (SN) and supermassive black holes (SMBH) feedback. The SN and SMBH feedback models in TNG use an isotropic energy injection. Hence, any directionality is a natural result of subsequent hydrodynamical and gravitational interactions. The model for stellar feedback uses a kinetic wind approach where star-forming gas is stochastically ejected from galaxies by Type II supernovae \citep{Springel2003}. Galactic winds are isotropic and the ejected wind particles are decoupled from surrounding gas in the star-forming environment until the density or time reaches a threshold, at which point they hydrodynamically recouple and deposit their mass, momentum, and energy \citep[see][for details]{Pillepich2018a}. The TNG stellar feedback model produces high mass loading galactic-scale outflows which are highly directional and metal-enriched \citep{Nelson2019a,Peroux2020,Ramesh2023a}.

In TNG, supermassive black holes are seeded in haloes exceeding a total mass of $\sim 7 \times 10^{10} M_\odot$. SMBHs then grow by merging with other black holes and accreting gas at the Eddington-limited Bondi rate. The mode in which the active galactic nucleus (AGN) ejects energy into its surroundings is dictated by this accretion rate. At low-accretion rates relative to the Eddington limit, kinetic energy is stochastically injected into neighbouring cells after enough energy is accumulated. On the other hand, thermal energy heats the surrounding gas at high-accretion rates proportional to the accreted mass. The thermal feedback mode typically dominates for SMBH masses $\lesssim 10^8$ M$_\odot$, and the kinetic mode for high-mass black holes \citep{Weinberger2017}. Feedback from SMBHs in the TNG model is the physical mechanism of galaxy quenching, ejecting gas from halo centers \citep{Truong2020} out to the scale of the closure radius \citep{Ayromlou2023}, while heating gaseous haloes and thus preventing future cooling of the CGM \citep{Zinger2020}.

\subsection{TNG50 Galaxy sample}

We identify haloes and subhaloes in the simulation using the \textsc{subfind} algorithm \citep{Springel2001}. The central subhalo is defined as the one found at the gravitational potential minimum of a friends-of-friends  halo \citep[FoF;][]{Davis1985} and the associated baryonic component is termed the central galaxy. Baryonic components of all other subhaloes associated with the FoF halo are dubbed satellites. 

We focus on galaxies at $z = 0.5$ with stellar masses ($M_*$) ranging from $\log(M_*/M_\odot) = 8.0$ to $11.0$ in order to match the properties of galaxies from the MUSE-ALMA Haloes survey \citep{Peroux2022, Karki2023}. The survey targets 32 high column density ($\logNHIunit > 18.0$) \ion{H}{i} absorbers and finds 79 galaxies within $\pm 500$ \kms\ of the absorbers at impact parameters ranging from 5 to 250 kpc \citep{Weng2023a}. 

We consider galaxies that are the central galaxies of their halo and measure their stellar masses within twice the stellar half mass radius. For the four stellar mass bins centred around $[10^{8.0}, 10^{9.0}, 10^{10.0}, 10^{11.0}]$\,$\rm{M_\odot}$, we select [100, 100, 50, 20] galaxies at random within bins of $\pm 0.3$\,dex. We choose the number of central galaxies in each mass bin such that the number of `sightlines' passing within the virial radius of the galaxies in each bin is approximately equal. The physical extent of the mocks along the plane perpendicular to the projection is the minimum of [400 pkpc, $2R_{\rm vir}$]. The limit of 400 pkpc is set by the physical scales covered in a single VLT/MUSE \citep{Bacon2010} pointing at $z \approx 0.5$. 

\subsection{Mock absorption columns}

The TNG simulations track the global neutral hydrogen content of gas cells. 
In order to split the gas between atomic or molecular hydrogen fractions individually, we adopt the molecular hydrogen fraction model of \citet{Gnedin2011} to estimate the H$_2$ fraction \citep[following][]{Popping2019}. 
We then obtain the \ion{H}{i} mass by subtracting the H$_2$ gas mass from the total neutral hydrogen content. 
The masses of all metallic ions including \ion{Mg}{ii}, \ion{C}{iv} and \ion{O}{vi} ions are computed in post-processing using v17.00 of the \textsc{cloudy} \citep{Ferland2017} code, assuming collisional and photoionization equilibrium in the presence of the meta-galactic UV background (UVB) from the 2011 update of \citet{Faucher2009} \citep[following][]{Nelson2020, Nelson2021}.

Our separation of galaxies into centrals versus satellites allows us to flag each gas cell based on whether it is gravitationally bound to the central, a satellite of the central or another FoF halo along the line of sight. 
Gas cells not bound to any halo (largely gas in the intergalactic medium) are also flagged. 
We henceforth refer to this family of flags as the set of (gravitational) \textit{origin} labels. 

Additionally, we separately categorise each gas cell as outflowing, inflowing or quasi-static using the cell's radial velocity ($v_r$) with respect to the central galaxy. 
{Hence, to be classified by one of these three flags, it is a pre-requisite that the gas cell is gravitationally bound to the central subhalo (designated as `central' in the origin labels).}
We calculate the radial velocity of each gas cell using the scalar product of the gas cell position relative to the galaxy centre with the velocity in the frame of reference of the subhalo excluding the Hubble flow. 
A radial velocity cutoff of $20$ \kms\ is used to identify inflowing and outflowing cells, that is, cells with $v_r > 20$ \kms\ are  outflowing and cells with $v_r < -20$ \kms\ are inflowing. 
Cells found at velocities $-20 < v_r < 20$ \kms\ are considered to be in quasi-hydrostatic equilibrium {and encompasses gas that is static or rotationally-dominated}. 
{We also include satellite galaxies in this set of flags so all gas cells within the central subhalo are included. }
Henceforth, this set of flags are the \textit{gas flow} labels. 
We consider other choices of the radial velocity boundary and the implications in section \ref{sec:flows}. 

\begin{figure*}
    \includegraphics[width=1.0\textwidth]{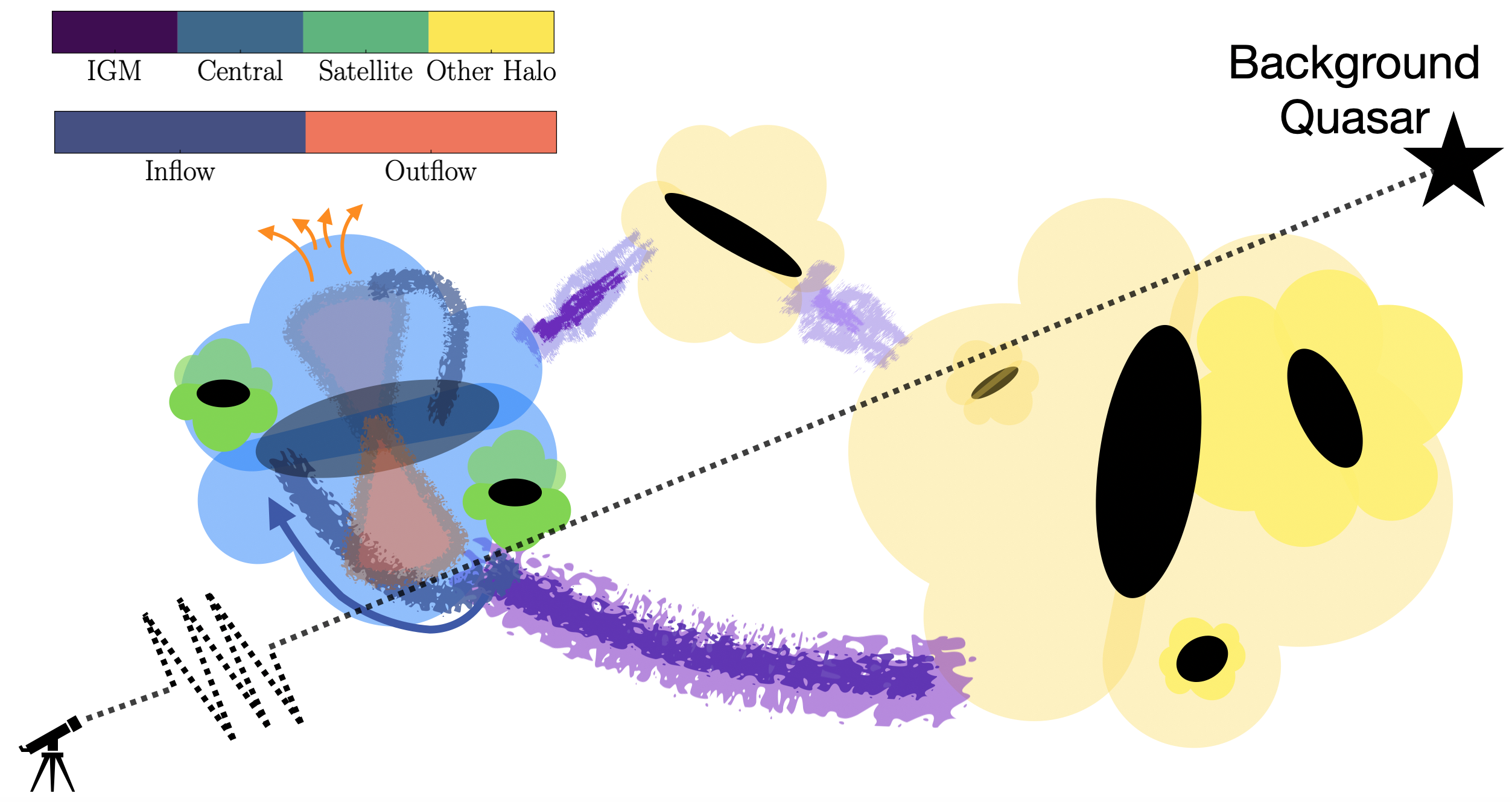}
    \caption{A cartoon view of intervening absorption lines towards background quasars intersecting galaxy haloes. Absorbers can have various origins in the circum- and intergalactic media. 
    They trace gas flows such as outflows (orange) and accretion (dark blue) as well as satellites (green). 
    There is also the possibility of the sightline intersecting other galaxy haloes along the line of sight (yellow) or filaments in the intergalactic medium (purple). 
    The gas can be found at a large range of impact parameters and velocities with respect to any given galaxy and the diversity of possible origins mean that connecting absorbers with their host galaxies requires a statistical approach. 
    We use the TNG50 simulation to quantify the contribution of the various physical origins presented, and how these contributions vary as a function of observable including the impact parameter, line of sight velocity, metallicity and azimuthal angle.}
    \label{fig:Sec2Cartoon}
\end{figure*}

We show a cartoon depicting the two sets of labels (origin and gas flow) in \autoref{fig:Sec2Cartoon}. 
Within the central galaxy's halo, they may trace gas accretion (bluish purple), outflows (orange) or satellite galaxies (green). 
However, absorbers may also intersect other galaxy haloes in front of or behind the central galaxy and these haloes are depicted in yellow. 
Finally, there is the chance that absorbers trace overdense gas in filamentary structures (purple). 
The various, intermixed possibilities highlight that relating the gas observed in absorption with its physical origin requires a statistical approach. 

To determine the column density of the \ion{H}{i} and other ions, we project the gas along the line of sight around each galaxy onto a large grid with pixel size $1\times1$ pkpc$^2$, through a line of sight depth of $\pm500$ \kms\ relative to the systemic velocity of the galaxy. 
{
We test pixel sizes with side length 0.5, 1 and 2 pkpc, finding only marginal differences in our forthcoming results. 
This is consistent with \citet{Szakacs2022}, where the H$_2$ and \ion{H}{i} column density distribution functions in TNG100 are found to differ only at column densities $\logNHIunit > 22$ when comparing pixel sizes with side length 150 pc and 1 kpc. 
These regions are limited to the centre of galaxies that comprise a small gas covering fraction when compared to the CGM. 
Hence, we adopt a 1 pkpc$^2$ pixel to match the median CGM resolution of $\approx1$ pkpc in TNG50 at $R_{\rm vir} > 0.1$ which increases for larger radii. 
}

The $\pm500$ \kms\ line of sight depth includes the contribution from the Hubble flow and we choose $\pm500$ \kms\ to be consistent with surveys of absorber-galaxy pairs \citep[e.g.][]{Dutta2020, Weng2023a}. 
We use the standard cubic-spline deposition method \citep[following][]{Nelson2016} to spatially distribute the masses of each gas component. 
Every pixel in this map corresponds to a single, observable `sightline'. 
For each pixel, we calculate the mass contribution from cells belonging to the origin and gas flow flags mentioned earlier: [IGM, central, satellite and other halo] (origin) and [inflow, quasi-static, outflow and satellite] (gas flow). 
We assign a flag to each pixel identifying which component contributes the most mass for that given sightline and then we determine the gas column density using the chosen flag only. 
The orientations of the central galaxies are random as we use three projections of a given halo that correspond to the fixed $x$-, $y$- and $z$-axes of the simulation volume. 

In addition to the column density, we calculate other quantities measurable in observations such as the mass-weighted (using \ion{H}{i}) metallicity of the gas, line of sight velocity difference between absorber and central galaxy ($\Delta v_{\rm LOS}$), two-dimensional projected distance from the galaxy centre (impact parameter, $b$) and azimuthal angle between the galaxy's major axis and absorber ($\Phi$).
We also include mass-weighted quantities such as the gas temperature, distance along the projection axis ($d_{\rm LOS}$) and distance from the galaxy centre (3D distance) which are not directly measurable in observations. 
These calculations are determined using only the cells that belong to a given flag. 
Ultimately, we generate a catalogue of absorbers with the properties discussed above that are labelled by two sets of flags that inform us of the origin and gas flow. 

\begin{figure*}
    \includegraphics[width=0.9\textwidth]{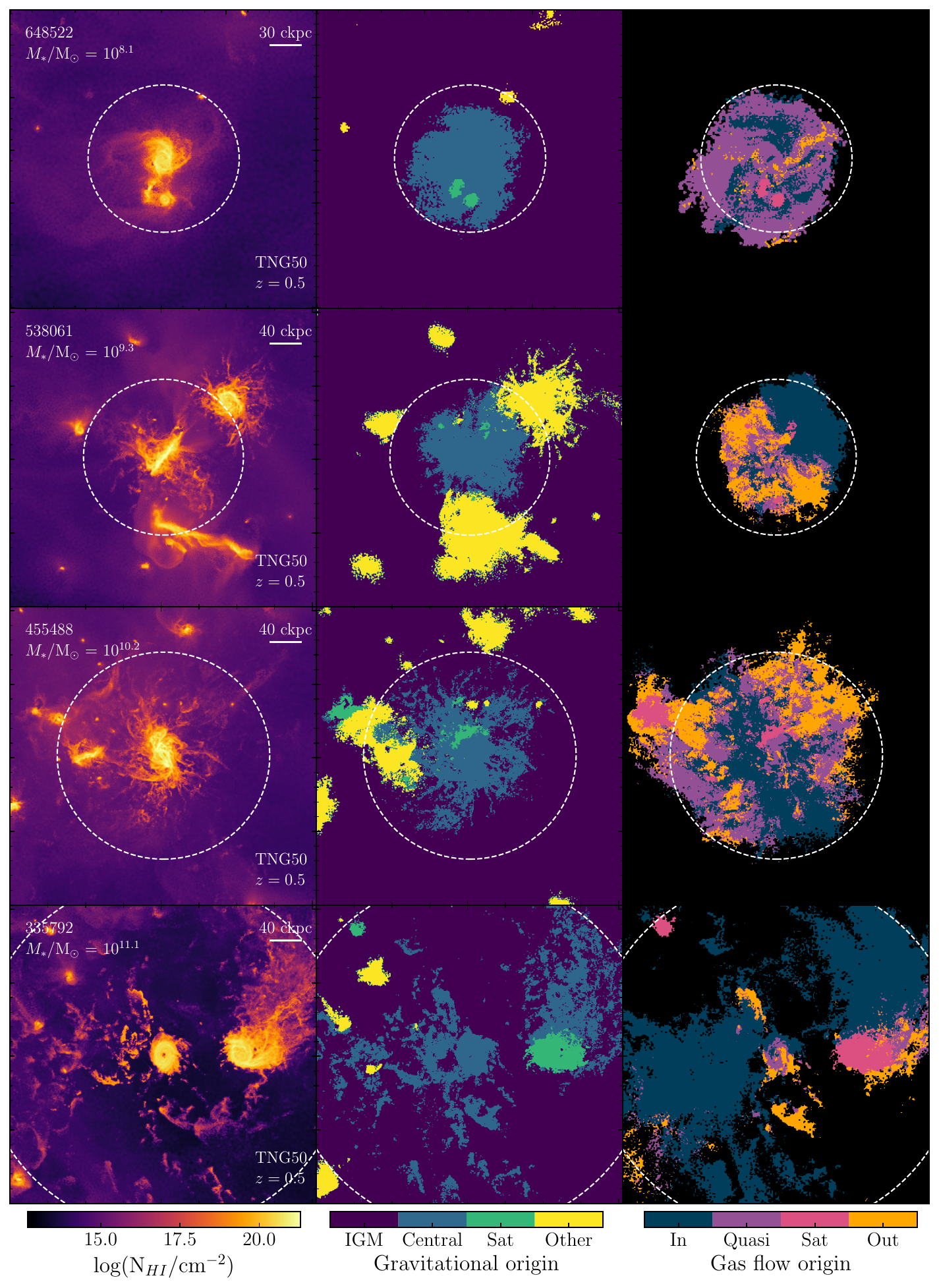}
    \caption{A visualisation of $z = 0.5$ galaxy haloes in the TNG50 sample of galaxies. 
    In the left column, we show the projected \ion{H}{i} column density maps of four galaxies belonging to different stellar mass bins from top to bottom. 
    The plots are also accompanied by the Subhalo index and a bar indicating the physical scale of the image. 
    The central and rightmost columns are the same galaxy shown in the left column but pixels are now coloured by the flag that dominates the \ion{H}{I} mass in projection. 
    {The black pixels in the right column signify that none of the gas flow labels meet the column density threshold of $\logNHIunit > 13.0$ in those regions. }
    The colours used directly correspond to the labels used in \autoref{fig:Sec2Cartoon}.
    We show the virial radius of these haloes with the dashed white circle. }
    \label{fig:Sec2TNGplot}
\end{figure*}

In \autoref{fig:Sec2TNGplot}, we display four randomly selected galaxies in the sample that span the entire range of stellar masses from roughly 10$^{8.0}$ to 10$^{11.0}$ M$_\odot$. 
The left column of plots are \ion{H}{i} column density maps where the subhalo identification, stellar mass and physical scale are also given. 
The central and rightmost plots separately colour each pixel by the dominant mass contribution for that sightline using the origin and gas flow flags. 
We keep the colour schemes for the two sets of flags consistent throughout the remainder of this work. 
The dashed white circle marks the virial radius of the central galaxy. 
From the first and second columns, we find that the dense \ion{H}{i} gas typically arises from the centre of centrals, satellites or other haloes, while lower \logNHI sightlines trace the intergalactic medium. 
Moreover, we see that satellites and other coincident galaxy haloes can dominate the projected \ion{H}{i} mass for a sightline even at small impact parameters from the central galaxy (most prominently seen in the second and third rows). 
Using the gas flow flags, we show that gas in the CGM has a rich structure in radial velocity. 
For the central galaxies depicted at larger inclinations (second and third rows), outflows appear to be preferentially directed along the minor axes. 
Gas accretion can be seen directed along the major axes but can arise from the central galaxy stripping gas from another galaxy (second row) or the accretion of gas clouds that co-rotate with the disk (bottom row). 
These velocity structures are washed out when the galaxy appears more face-on (top row), where much of the gas is quasi-static. 

{We note that there are subtleties in the categorisation of pixels using the origin and gas flow depicted in Figures \ref{fig:Sec2Cartoon} and \ref{fig:Sec2TNGplot}. 
First, we separate gas inflows from satellites that may be accreting onto galaxies. 
Components of accreting satellites that have been stripped and are infalling onto the central galaxy will be considered `inflows', whereas gas that is still more tightly bound to the satellite will be labelled as `satellite'. 
Similarly, large-scale streams from the IGM, as seen in the rightmost panels for $10^8$ and $10^{11}$ M$_\odot$ central galaxies, will be considered inflows only if they are gravitationally bound to the central galaxy. 
In the middle column, yellow and purple dominates the regions beyond the virial radius because other haloes and the intergalactic medium contribute the most \ion{H}{i} for these sightlines. 
As we do not consider the second halo term and IGM in the rightmost column, the black pixels correspond to regions where none of the four gas flow labels [inflow, quasi-static, outflow and satellite] meet a column density threshold of $\logNHIunit > 13.0$ (see \autoref{fig:Sec2TNGplot}). 
It is also for this reason that the central galaxy and satellites appear more extended in the right column of that same figure (e.g. satellites in the top left corner of the bottom two rows). 
Pixels previously assigned to the `IGM' or `other halo' origin labels may change into the most dominant of the four gas flow labels, assuming the column density threshold is met. 
}

\section{Where is the gas along the line of sight?}
\label{sec:where}
In observational studies of absorber-galaxy systems, the impact parameter and line of sight velocity are two reliably-determined measurables that inform us on the true distance between absorbers and galaxies, and, by extension, determine how absorbers are related to surrounding galaxies. While the impact parameter is a measure of the physical distance in the plane of the sky, the velocity difference ($\Delta v_{\rm LOS}$) does not directly correlate with the distance along the line of sight ($d_{\rm LOS}$). An absorber's line of sight velocity is a combination of the gas flow being traced, viewing angle of the galaxy and the Hubble flow. Hence, it is important to evaluate the position of the absorbers with $\Delta v_{\rm LOS}$ within 300 to 1000 \kms \ of the galaxy systemic redshift \citep[typical values used in recent surveys:][]{Schroetter2016, Peroux2022, Galbiati2023} may be beyond the virial radius. In this section, we examine the relation between $\Delta v_{\rm LOS}$ and line of sight distance, and how the relation evolves as a function of the gas origin and motion (e.g. absorbers tracing the IGM, accretion and outflows). Additionally, we estimate the ranges in $\Delta v_{\rm LOS}$ where the majority of the CGM gas mass lies. 

\begin{figure*}
    \includegraphics[width=1.0\textwidth]{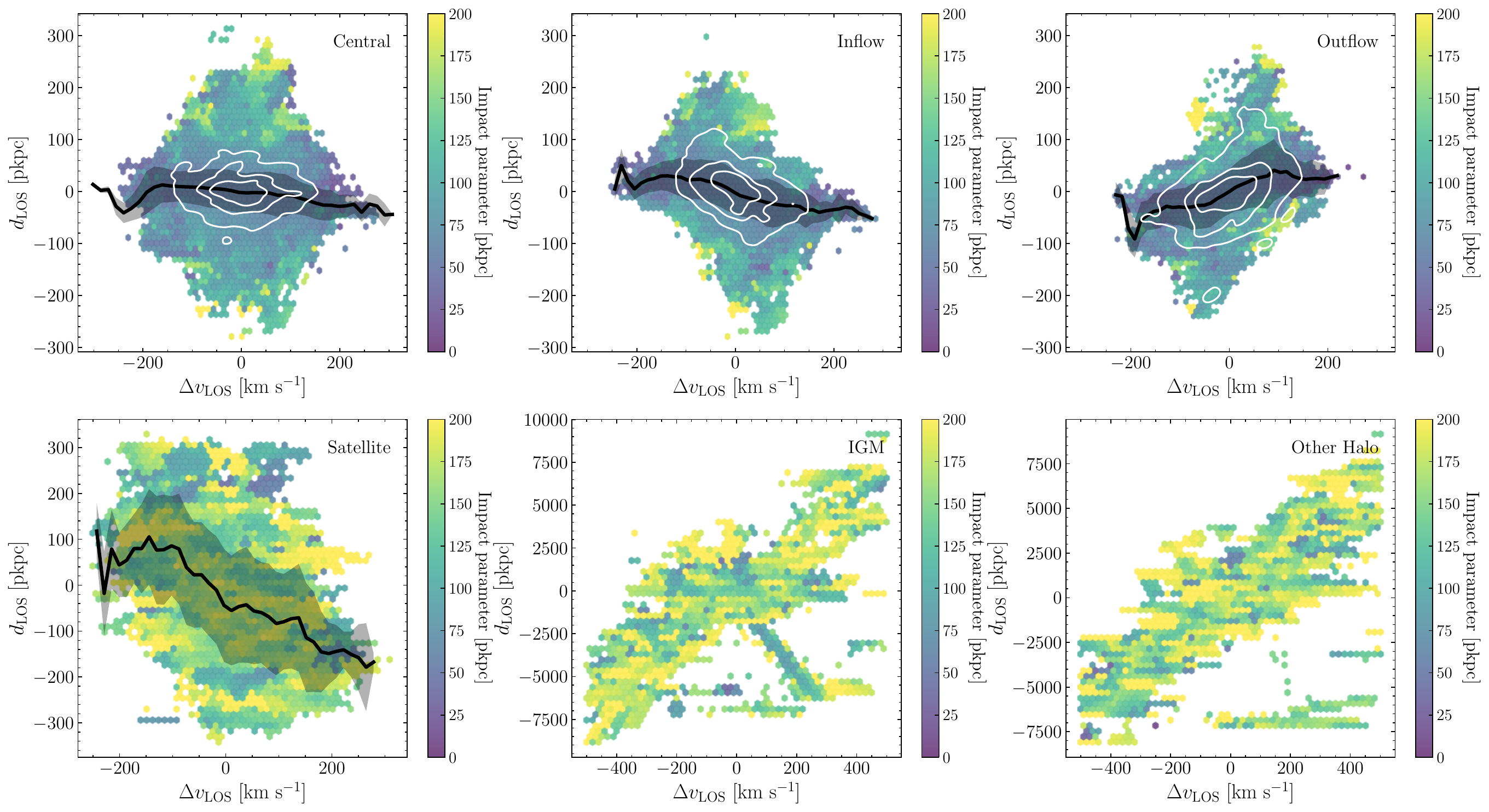}
    \caption{A comparison of the line of sight distance with the line of sight velocity for mock absorbers in TNG50 at $z = 0.5$. 
    Positive (negative) line of sight distances are behind (in front of) the central galaxy away from (towards) the observer. 
    Similarly, positive (negative) line of sight velocities are directed away from (towards) the observer. 
    We consider \ion{H}{i} absorbers with column densities $16.0 < \logNHIunit < 17.2$ (partial Lyman-limit systems) associated with $\log(M_*/M_\odot) \approx 10.0$ central galaxies. 
    Absorbers are separated by their origin or gas flow which are given in the top right corner of each plot. 
    Each hexbin is coloured by the median impact parameter of all absorbers found within that hexbin. 
    We plot the mean $d_{\rm LOS}$ in 40 bins with width $\sim$10 \kms \ for the `Central', `Inflow', `Outflow' and `Satellite' panels as a thick black line and the shaded region represents the standard deviation. 
    In general, close separations in line of sight velocity can correspond to significant separations in line of sight distance, even when an absorber is tracing the central galaxy. 
    Using a kernel density estimator, the white contours on the top row show where the absorbers are concentrated with contour levels of [0.25, 0.50, 0.75] times the total number of absorbers. 
    We find that 75 per cent of the \ion{H}{i} mass traced by pLLS is enclosed within $|\Delta v| < 150$ \kms\ of the 10$^{10}$ M$_\odot$ central galaxies. 
    This $\Delta v$ limit varies between 70 and 250 \kms\ for the stellar mass range in our sample but does not change significantly with column density. }
    \label{fig:Sec3dLOS}
\end{figure*}

In \autoref{fig:Sec3dLOS}, we show the relationship between the line of sight distance and line of sight velocity for absorbers associated with a stacked sample of central galaxies with stellar mass $\log(M_*/M_\odot) = 10.0$ at $z = 0.5$. This particular figure is limited to partial Lyman limit systems (pLLS; $16.0 < \logNHIunit < 17.2$) associated with 10$^{10}$ M$_\odot$ central galaxies. Because this choice is arbitrary, we discuss the effects of changing column density and stellar mass. We separate absorbers by their respective origin or gas flow. The top-left panel includes all absorbers associated to the central galaxy. The panels labelled as `Inflow' and `Outflow' in the top row are decompositions of the top-left plot (we have not included `quasi-static' gas with radial velocities $|v_r| < 20$ \kms). We display the properties of absorbers that trace gas in satellites, the intergalactic medium or another halo along the line of sight in the bottom row respectively from left to right. 
The colour of each hexbin is the median impact parameter of all absorbers found in the bin. 

For gas associated with the central galaxy and its satellites, we find a wide array of line of sight distances for any given line of sight velocity, with values varying by up to $\sim$\,$500$ pkpc at fixed $\Delta v_{\rm LOS}$. 
One cause for this is the degeneracy between the physical location of the absorbing gas and its kinematics. 
Redshifted absorbers can trace gas accreting onto the galaxy from behind or gas ejected towards the observer. 
Similarly, blueshifted absorption lines can probe outflowing gas expelled away from the observer or gas infalling onto the galaxy from in front. 
This is evident from the black horizontal bars that display the average $d_{\rm LOS}$ associated with absorbers in 40 bins with width $\sim$ 10\kms in \autoref{fig:Sec3dLOS}. 
Gas at negative line of sight velocities trace both inflowing (outflowing) gas at positive (negative) distances down the sightline. 
For centrals, the black line is almost horizontal at $d_{\rm LOS} = 0$ kpc and there is roughly $50$ kpc of scatter in $d_{\rm LOS}$ at all $\Delta v_{\rm LOS}$. 
In general, inflows are indistinguishable from outflows by the velocity of the absorber relative to the galaxy alone \citep{Tumlinson2017} and we require other diagnostics to break the degeneracy (see Section \ref{sec:flows}).

We also find absorbers associated with satellite galaxies, the intergalactic medium and other galaxy haloes. 
In particular, \autoref{fig:Sec3dLOS} shows that partial Lyman-limit systems with $|\Delta v_{\rm LOS|} < 500$ \kms \ can trace gas in the IGM or intersect other haloes that are located more than several pMpc along the line of sight. 
While these absorbers are typically found at larger impact parameters, we still find that gas in the IGM and other haloes can mimic CGM gas of the central galaxy. We quantify the relative frequency of these absorbers in Section \ref{sec:sat}. 
The black horizontal bars indicating the median line of sight velocity for gas tracing satellites is similar to absorbers tracing accretion, highlighting that satellites are infalling onto the central and their average motion resembles gas accretion.  
In contrast, the almost linear relationship between $d_{\rm LOS}$ and $\Delta v_{\rm LOS}$ for gas in the IGM and other haloes is caused by the Hubble flow dominating at larger scales. We note that the range in LOS distances of the `IGM' and `Other Halo' panels is not the same as the other four. 

Nevertheless, it is clear from the inflow-outflow degeneracy combined with the contribution of absorbers from satellites, the IGM and other haloes that the line of sight velocity difference between absorber and galaxy is not a direct indicator of actual distance. By extension, a small velocity separation (e.g. $|\Delta v_{\rm LOS}| < 500$ \kms) does not imply an absorber is associated with a given galaxy. This result is in line with the findings of previous works using the EAGLE simulations \citep{Schaye2015, Crain2015} to analyse \ion{H}{i} around massive galaxies (M$_{200}$\,$\gtrsim$\,$10^{12}~\rm{M_\odot}$) at $z \sim 2$$-$3 \citep{Rahmati2015}, and \ion{Mg}{ii} and \ion{O}{vi} around stellar mass $10^9$ to 10$^{11}$ M$_\odot$ galaxies at $z \approx 0.3$ \citep{Ho2020, Ho2021}. More recently, the FOGGIE simluations, find that structures of varying temperatures, metallicities and densities in the CGM that span $> 100$ kpc can be tightly constrained in velocity space \citep{Peeples2019}. We similarly show here that the velocity restrictions we apply in observations lead to a larger LOS path length compared to the virial radius of galaxies. In addition, when multiple absorption components are separated in velocity (from tens to hundreds of \kms) down a single sightline in observations, our results emphasise that even these smaller velocity differences can lead to different spatial locations. Studies find discrepancies larger than 2 dex in metallicity for distinct absorption components \citep[e.g.][]{Zahedy2019, Lehner2022} and we suggest that these separate components may arise from different physical origins such as the intergalactic medium. 

We are able to give a simple prescription for an appropriate line of sight velocity limit that encompasses most of the \ion{H}{i} gas belonging to the CGM of the central using the distribution of absorbers in velocity space. 
Using a kernel density estimator with contour levels of [0.25, 0.50, 0.75] in \autoref{fig:Sec3dLOS}, we find that 75 per cent of LLS absorbers associated with the central are found within $\pm 150$ \kms of galaxies with  stellar mass 10$^{10}$ M$_\odot$. 
In addition, we find that our prescription of $\pm 150$\ \kms\ varies marginally for absorbers of \ion{H}{i} column densities larger than 10$^{17.2}$ atoms cm$^{-2}$. 
However, this value is strongly dependent on the stellar (and more importantly, the halo) mass and decreases to $\pm 70$ (120) \kms\ for $M_* \approx 10^{8}$ ($10^{9}$) M$_\odot$ galaxies and is $\pm 250$ \kms\ for the $M_* \approx 10^{11}$ M$_\odot$ galaxies in our sample. 
Fundamentally, the increasing virial velocities for more massive haloes drives the variation in $v_{\rm LOS}$ values. 
However, this relation is not linear, with $\Delta v_{\rm LOS} > V_{200}$ {(halo circular velocity)} for smaller haloes, and vice versa, for larger haloes. 
We also find these values are roughly consistent with the findings of \citet{Peeples2019}, where the strongest absorption is confined to be within $\pm 200$ \kms\ of a $M_* \approx 10^{10}$ M$_\odot$ galaxy at $z = 2$. 
We refrain from a more direct comparison with FOGGIE because their cosmological zoom simulations are at a different redshift and do not include contributions from the IGM or other haloes. 
These $\Delta v_{\rm LOS}$ estimates serve as useful indicators of whether an \ion{H}{i} absorber is within the halo of a nearby galaxy. 

\section{The physical origin of absorbers}
\label{sec:sat}
While the limits in $\Delta v_{\rm LOS}$ provided in the previous section are practically useful for observational studies of absorber-galaxy systems, it is also important to quantify the likelihood of absorbers originating from the central galaxy as opposed to satellites, other galaxy haloes along the line of sight or the intergalactic medium. Here, we explore how the fraction of sightlines that trace gas outside the central subhalo changes as a function of impact parameter and \ion{H}{i} column density, two observables that are readily available. In addition, we specifically study the contribution of lower mass satellites that are below the current sensitivity limit of many observations ($M_* \lesssim 10^8$ M$_\odot$). 

\subsection{What fraction of \ion{H}{i} arise from the central halo?}
We have shown that the combination of gas peculiar velocities and the Hubble flow lead to absorbers of various origins masquerading as belonging to the central galaxy. Here, we quantify the fractional contribution of the various gas origins as a function of the impact parameter. We use the impact parameter in particular because typically, the galaxy at lowest impact parameter is considered to be associated with the absorber in observational surveys \citep[e.g.][]{Langan2023, Weng2023a}. Likewise, we consider all gas within $\pm 500$ \kms\ of the central galaxy to mimic a typical line of sight velocity cut in observations \citep{Dutta2020, Dutta2021, Hamanowicz2020, Galbiati2023, Weng2023a}.  

\begin{figure*}
    \includegraphics[width=1.0\textwidth]{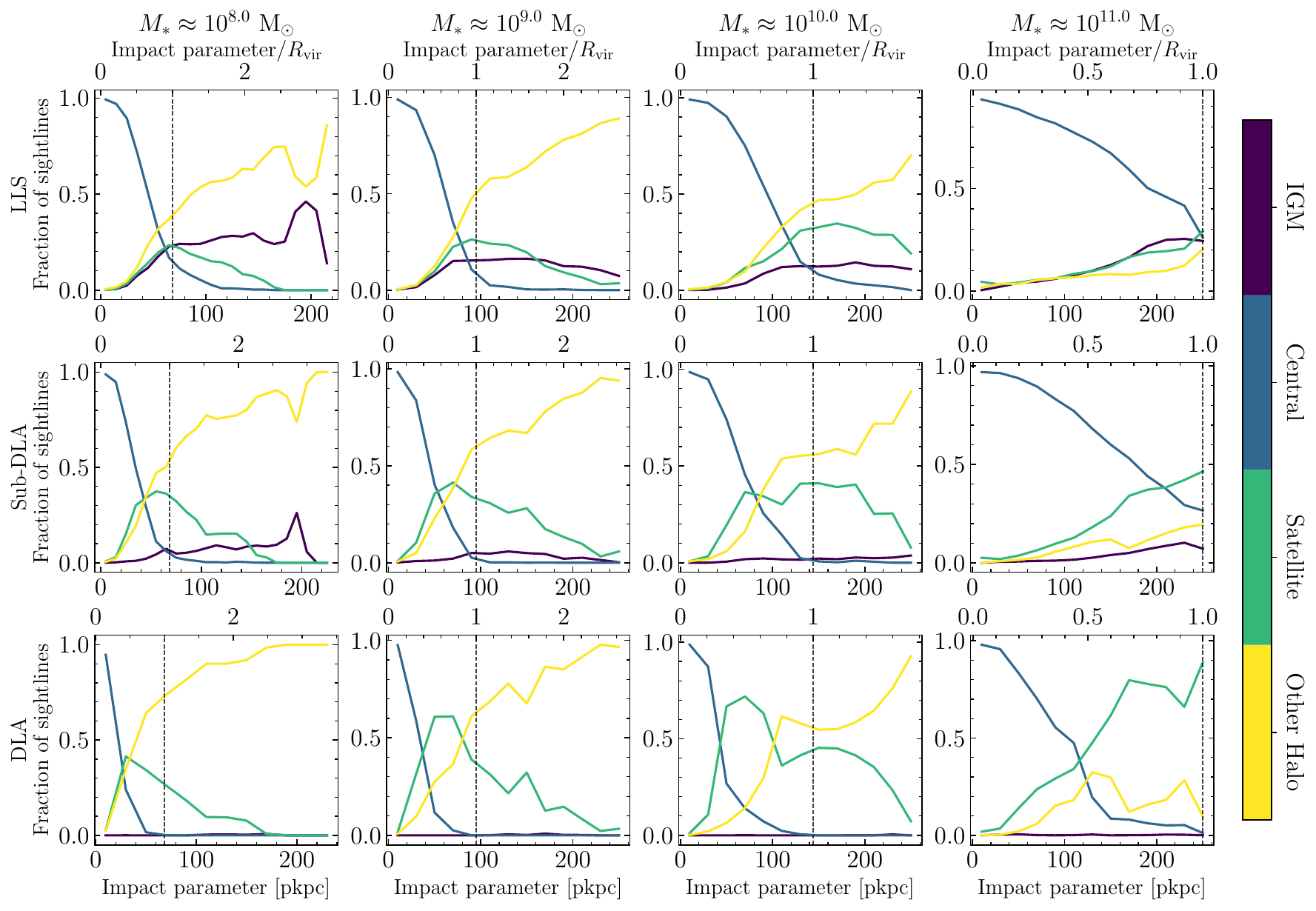}
    \caption{The fraction of \ion{H}{i} absorbers that arise from gas in the IGM, central, satellites or other haloes as a function of the impact parameter (normalized by $R_{\rm vir}$ in the top axis) at $z = 0.5$. 
    A vertical dashed black line indicates the virial radius. 
    Here, we consider four stellar mass bins ranging from $\log(M_*/M_\odot) = 8.0$ to $11.0$ in one dex intervals and three column density ranges: $17.2 < \logNHIunit < 19.0$ (LLS), $19.0 < \logNHIunit < 20.3$ (sub-DLAs) and bottom; $\logNHIunit > 20.3$ (DLAs).
    The central galaxy dominates at impact parameters up to $\approx0.5$ $R_{\rm vir}$, though this limit decreases for higher column densities. 
    At roughly $0.5R_{\rm vir} < b < R_{\rm vir}$, satellite galaxies dominate the fraction of absorbers and beyond the virial radius, other haloes along the line of sight contribute most to the absorber count. 
    There is only a marginal contribution from the intergalactic medium at these \ion{H}{i} column densities. 
    These results highlight that knowledge of the stellar mass of galaxies associated with quasar absorbers is key to relate the gas probed in absorption with its origin. 
    }
    \label{fig:Sec4SATfcover}
\end{figure*}

In \autoref{fig:Sec4SATfcover}, we show the fraction of absorbers that trace gas in the central (blue), a satellite of the central (green), another galaxy halo along the line of sight (yellow) and gas in the intergalactic medium (purple). We express the fraction as a function of impact parameter (normalized by the virial radius in the top $x$-axis) for each of the four origins. The twelve panels span four stellar mass bins in ascending order from left to right and three \ion{H}{i} column density ranges: LLS (top; $17.2 < \logNHIunit < 19.0$), sub-DLAs (centre; $19.0 < \logNHIunit < 20.3$) and DLAs (bottom; $\logNHIunit > 20.3$). We use bins of size 10 (20) kpc to calculate the fractions for the LLS and sub-DLA (DLA) column densities. 

We find that these fractions are strongly dependent on the stellar mass of the central galaxy. Absorbers associated with more massive central galaxies dominate out to larger physical impact parameters, but this is caused by more massive galaxies having a larger virial radius (see normalized $b/R_{\rm vir}$ on the top $x$-axis). 
Indeed, the slopes of the central curves when viewed as a function of the normalized impact parameter are almost identical across our four stellar mass bins. 
Moreover, for stellar masses $\log(M_*/M_\odot) \leq 10.0$, we often find a peak in the contribution from satellites before the ‘other halo’ term begins dominating. This is caused by the fact that satellites are limited to being gravitationally bound to the central. It is only for the 10$^{11}$ M$_\odot$ central galaxies where the virial radius is larger than the mocks (as seen in \autoref{fig:Sec2TNGplot}) that the contributions from satellite galaxies have not peaked. We note again that these size restrictions for the most massive galaxies in the plane perpendicular to projection are physically motivated by the scales covered by the field of view of VLT/MUSE at $z \sim 0.5$ which is the median redshift of the MUSE-ALMA Haloes survey \citep{Peroux2022}. Ultimately, the general trend we find, which is roughly independent of the mass and column density, is that the central galaxy dominates the covering fraction up to $b \approx 0.5R_{\rm vir}$, followed by satellite galaxies until the virial radius and then other galaxy haloes beyond $R_{\rm vir}$. 

There are also differences between the \ion{H}{i} column density bins (increasing column density from top to bottom rows). \autoref{fig:Sec4SATfcover} reveals a steeper decline in absorbers associated with centrals for higher \ion{H}{i} column densities, as these densities drop rapidly in the CGM towards larger impact parameters \citep[e.g][]{Ramesh2023a}. This is accompanied by a greater fractional contribution of satellites and other haloes at these impact parameters, as a result of sightlines intersecting the dense ISM of these objects. The curves are more irregular because strong absorbers are rarer, even with larger bin sizes. We also see that the contribution from gas in the intergalactic medium (purple) is minor for column densities $\logNHIunit \geq 19.0$, but increases in prevalence at lower N$_{\ion{H}{i}}$ and larger impact parameters, as expected since this gas is predominantly not \ion{H}{i} as a result of ionization by the UV background. 

\begin{figure*}
    \includegraphics[width=1.0\textwidth]{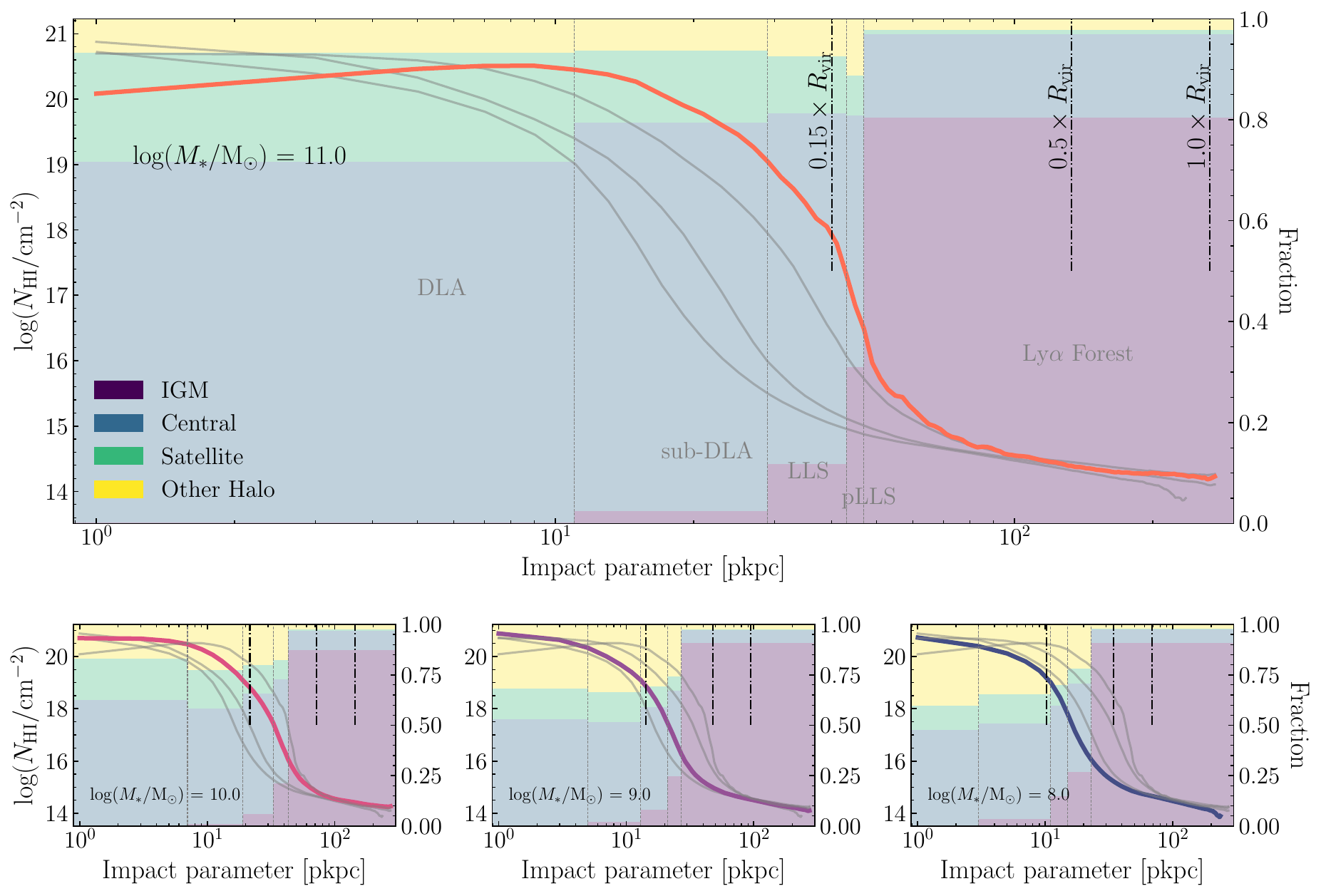}
    \caption{The median \ion{H}{i} column density of absorbers versus given impact parameter for galaxies with stellar masses covering $\log(M_*/M_\odot) = 8.0$ to $11.0$ at $z = 0.5$. We show the $N_{\ion{H}{i}}$-$b$ relation in the primary panel for $M_* = 10^{11}$ M$_\odot$ central galaxies with the coloured line. The faded grey lines represent the same relation but for the lower stellar mass bins found in the secondary panels. 
    The vertical dashed grey lines mark the impact parameters where the median $\logNHI$ transitions from damped Ly-$\alpha$ absorbers to partial Lyman-limit systems for $\log(M_*/M_\odot) = 11.0$ galaxies. The last impact parameter bin covers column densities $13.0 < \logNHIunit < 16.0$ (Ly-$\alpha$ forest). The coloured area in each bin is proportional to the fraction of absorbers (with \logNHI consistent with the respective bin) that belong to the categories of IGM, central, satellite and other halo. For example, the first bin that covers DLAs shows that most $\logNHIunit > 20.3$ absorbers will arise from the $M_* = 10^{11}$ M$_\odot$ central galaxy, with small contributions from satellites and other haloes along the line of sight. At lower column densities, the fraction of sightlines that begin to trace gas in the intergalactic medium increases and then the IGM dominates for absorbers at Ly-$\alpha$ forest column densities. 
    Similar to \autoref{fig:Sec4SATfcover}, we find that the contribution of satellites, other haloes and the IGM increases with the impact parameter. 
    Additionally, we notice that higher mass central galaxies account for a larger fraction of dense \ion{H}{i} absorbers ($\logNHIunit > 17.2$) but at lower $N_{\ion{H}{i}}$, the central fractions remain consistent across the four stellar mass bins. 
    Black dashed vertical lines mark impact parameters that are 0.15, 0.5 and 1 times the virial radius. 
    Despite the various potential contributions from satellites, interloping haloes or the intergalactic medium, the trend of decreasing \ion{H}{i} column density with impact parameter remains clear.}
    \label{fig:Sec4NHIb}
\end{figure*}

This becomes clearer in \autoref{fig:Sec4NHIb} where we plot the trend of \ion{H}{i} column density as a function of the impact parameter for the four stellar mass bins. 
The solid coloured line in the primary panel is the median $N_{\ion{H}{i}}$ value at a given impact parameter for $M_* = 10^{11}$ M$_\odot$ central galaxies. 
The faded grey curves represent the $N_{\ion{H}{i}}$-$b$ for the lower stellar mass bins which are shown and coloured in the secondary panels. 
Grey vertical lines mark the impact parameters where the median column density drops below the DLA, sub-DLA, LLS and pLLS thresholds. 
The coloured area is proportional to the fractional contribution (right $y$-axis) of the four origin flags for each range in column density. 
For instance, DLAs found towards our $M_* = 10^{11}$ M$_\odot$ central galaxies (primary panel) can be attributed 75 per cent of the time to the central, 15 per cent to satellites and the remainder to other galaxy haloes. 
This calculation is made using DLAs found across all impact parameters, as opposed to \autoref{fig:Sec4SATfcover} where the fractions are given as a function of $b$. 
The black dashed vertical lines mark 0.15, 0.5 and 1 times the virial radius to mark the inner and outer boundaries of the CGM. 

In all panels of \autoref{fig:Sec4NHIb}, we find an increase in the incidence of absorbers tracing the intergalactic medium as we move to larger impact parameters (and thus lower median \ion{H}{i} column densities).
This highlights that weaker absorbers have a high likelihood of originating from the IGM, with more than 30 per cent of \logNHI < 10$^{16}$ atoms cm$^{-2}$ absorbers tracing gas outside of haloes. 
We also see a larger fraction of absorbers belong to the central across all column densities for the more massive haloes. 
Crucially, we note that this signal is caused by our restrictions in the mock size perpendicular to the plane of projection which is intended to mimic observational surveys that have fixed observed size. 
The less massive central galaxies will have a larger portion of their maps outside the virial radius and hence, have a higher incidence of absorbers arising from other haloes or the intergalactic medium. 

We also see in \autoref{fig:Sec4NHIb} that the median \ion{H}{i} column density is larger at all impact parameters $\gtrsim 10$ kpc for the more massive central galaxy. 
At smaller impact parameters, this trend does not hold because of the supermassive black hole at the centre of the most massive galaxies ejecting gas via the kinetic feedback mode \citep{Zinger2020}. 
In addition, we find a scatter in $N_{\ion{H}{i}}$ up to 3 dex across the range of impact parameters. 
This scatter can be attributed to the intrinsic inhomogeneity of gas in the circumgalactic medium, but our results also emphasise that we may be intersecting gas clouds outside galaxy haloes in the IGM. 
Despite the various potential contributions from satellites, interloping haloes or the intergalactic medium, the trend of decreasing \ion{H}{i} column density with impact parameter remains clear \citep{Lehner2013, Nelson2020, vandevoort2021, Berg2023, Weng2023a}.

\subsection{What fraction of absorbers intersect low-mass satellites?}
The Milky Way halo contains tens of satellite galaxies \citep{McConnachie2012} and we expect absorption lines to occasionally intersect these smaller haloes found in the CGM of more massive galaxies. 
The probability of sightlines intersecting satellites, particularly faint ones, is of interest to observers as strong absorbers occasionally do not have galaxy counterparts down to some magnitude ($R \approx 25$ mag) and/or star-formation rate (SFR) limit ($SFR \approx 0.1$) \citep[e.g.][]{Fumagalli2015, Berg2023, Weng2023a}. 
Here, we estimate the gas mass contribution from low-mass satellite galaxies in the TNG50 simulation. 
This is motivated by the shallower completeness expected for galaxies below some stellar mass limit in observational surveys. 

\begin{figure*}
    \includegraphics[width=1.0\textwidth]{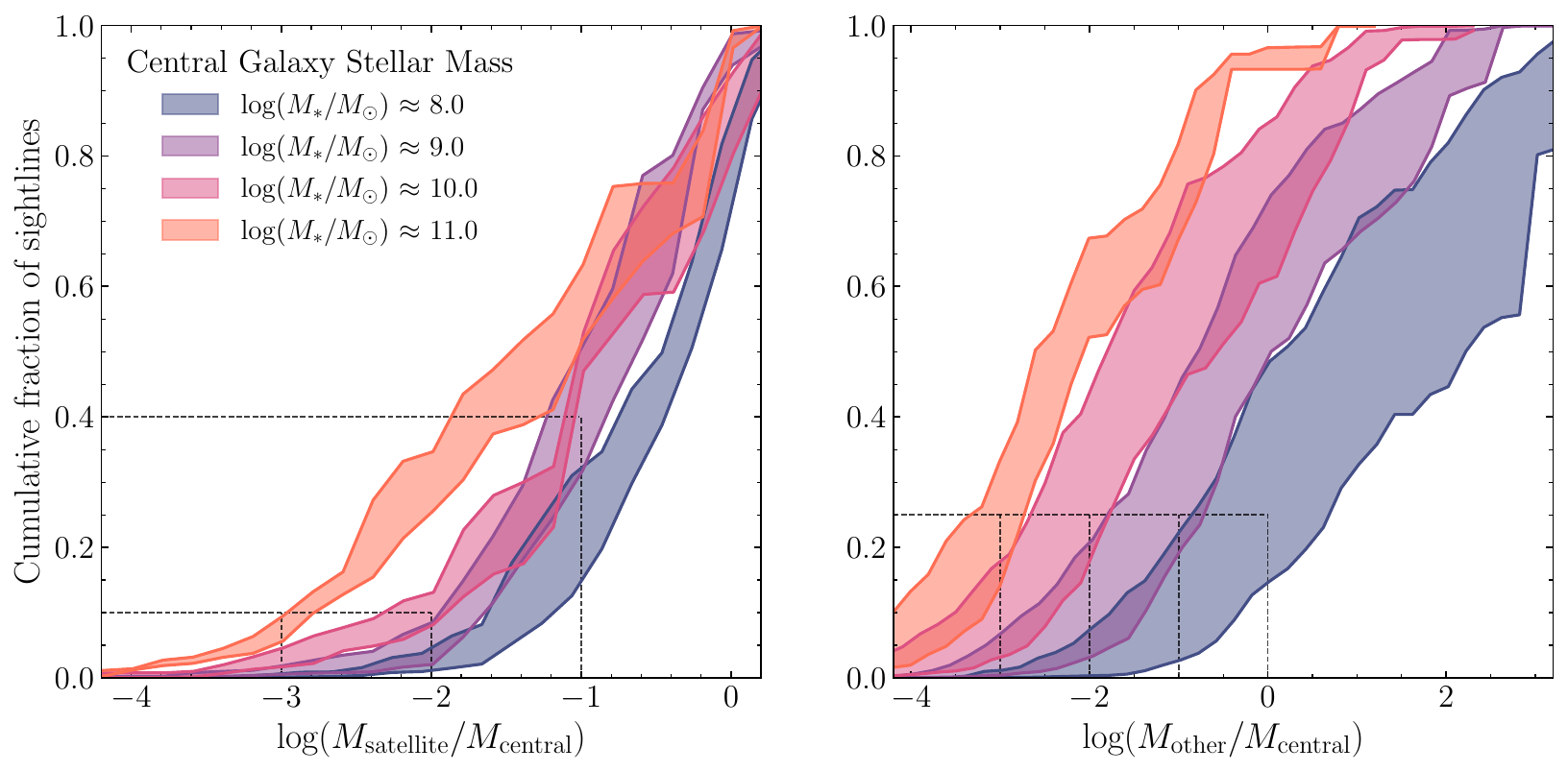}
    \caption{The contribution of satellites (left) and other haloes (right) with varying masses to the fraction of sightlines that intersect satellites. 
    The four colours correspond to the four stellar mass bins for the central galaxy. 
    We performed this calculation using five column density bins, from Ly-$\alpha$ forest column densities ($13.0 \leq \logNHIunit \leq 16.0$) to DLAs. 
    The shaded region between the solid lines for each stellar mass bin reflects the maximum discrepancy (typically $\sim$0.1) in the fraction found between the five column density bins. 
    We use stellar mass bins of size $\log(M_*/M_\odot) = 0.2$. 
    On the left, we find that galaxies with masses $M_* < 10^8$ M$_\odot$, which are possibly missed in current observational surveys, comprise 10 per cent of sightlines that intersect satellites for central galaxies with stellar masses 10$^{10}$ and 10$^{11}$ M$_\odot$. 
    This value increases to 40 per cent for 10$^{9}$ M$_\odot$ central galaxies. 
    On the right, secondary haloes belonging to galaxies with $M_* < 10^8$ M$_\odot$ contribute to 25 per cent of sightlines that intersect another halo across all central masses. 
    The dashed vertical and horizontal lines help guide the eye to the numbers given above. 
    We emphasise here that low-mass galaxies account for a significant portion of the \ion{H}{i} covering fraction. 
    }    
    \label{fig:Sec4LowMassSat}
\end{figure*}

For a given TNG50 mock with satellites and other haloes, each gas cell is assigned a subhalo ID. 
We calculate the subhalo ID that dominates the HI mass for each pixel in our 2D projection down the line of sight of the mock. 
By crossmatching this list of IDs with sightlines that have been assigned a ‘satellite’ or 'other halo' flag, we then compute the fractional contribution of galaxies with varying stellar masses to the total number of sightlines that are dominated by gas in satellites or a secondary halo along the sightline. 
This is depicted in \autoref{fig:Sec4LowMassSat} where the four colours correspond to the four stellar mass bins in our sample. 
We show the cumulative contribution from satellites (left) and other haloes (right) with stellar masses ($M_{\rm satellite}$) that begin at 10$^{-4}$ times the mass of the central galaxy ($M_{\rm central}$). 
The shaded region reflects the scatter when considering different column density bins, from $13.0 < \logNHIunit < 16.0$ to $\logNHIunit > 20.3$ systems. 

For $M_* = 10^{10}$ and $10^{11}$ M$_\odot$ central galaxies, $M_* < 10^8$ M$_\odot$ satellite galaxies only contribute roughly ten per cent to the total number of absorbers that intersect any satellite (see dashed vertical and horizontal lines on the left plot). 
This fraction increases to $\approx$40 per cent of sightlines for a central galaxy mass of $10^9$ M$_\odot$. 
For other haloes along the line of sight, the contribution from $M_* < 10^8$ M$_\odot$ galaxies remains roughly constant at 25 per cent (right). 
There is a larger scatter of roughly 0.2 between the different column density bins, but the upper and lower bounds are not driven by any particular column density. 
This effect highlights that even low-mass galaxies in the CGM of massive galaxies and along the line of sight contribute to absorption and might remain undetected in observational surveys of absorbers, particularly at high redshift. 

There are occasional cases in observational studies of galaxy counterparts to absorbers where no galaxy is detected near the QSO sightline down to some limit in stellar mass or star-formation rate \citep{Fumagalli2015, Berg2023, Weng2023a}. 
At $z>2$ in surveys of Ly-$\alpha$ emitters (LAEs) associated with \ion{H}{i} absorbers, there are also systems where LAEs are found at impact parameters $> 50$ kpc from DLAs \citep{Mackenzie2019, Lofthouse2023}. 
Given that the \ion{H}{i} is not expected to extend out to such large distances from simulations \citep{Stern2021}, the more plausible explanation is that we are tracing low-mass satellite galaxies near the QSO sightline. 
Figure \ref{fig:Sec4SATfcover} suggests that satellite galaxies at $z=0.5$ are responsible for up to 50 per cent of absorbers at impact parameters within the virial radius of the central but particularly at $0.5 < R_{\rm vir} < 1.0$. 
We then see in \autoref{fig:Sec4LowMassSat} that $M_* \leq 10^8$ M$_\odot$ satellites make up $\approx$10 (40) per cent of the total number of sightlines that trace satellites gravitationally bound to central galaxies with stellar masses 10$^{10}$ and 10$^{11}$ (10$^{9}$) M$_\odot$.
Hence, strong \ion{H}{i} absorbers without a nearby galaxy counterpart may simply be associated with objects below the sensitivity limit (typically $10^8$ M$_\odot$ at $z = 0.5$ for a one hour MUSE exposure). 
Such objects require deeper observations or larger telescopes such as the Extremely Large Telescope (ELT), Giant Magellan Telescope (GMT) or the Thirty Meter Telescope (TMT). 

\section{Gas flows in the CGM}
\label{sec:flows}
As depicted in Figure \ref{fig:Sec2Cartoon} and \ref{fig:Sec2TNGplot}, disentangling inflowing from outflowing gas in observations of the circumgalactic medium is not straightforward. 
Two commonly used indicators to observationally infer whether the gas is inflowing or outflowing are the azimuthal angle ($\Phi$) and metallicity ($Z$). 
In this section, we characterise how the inflowing to outflowing gas fraction changes as a function of $\Phi$ and study the origins of the metallicity anistropy between the major and minor axes of galaxies.  

\subsection{Using the azimuthal angle to identify gas flows}
In transverse absorption-line studies, the 2D projected azimuthal angle is often advocated as an indicator to distinguish outflows from inflows \citep[e.g.][]{Bouche2012, Kacprzak2012, Schroetter2016, Schroetter2019, Weng2023b} where accretion is assumed to align with the major axis ($\Phi = 0^\circ$) and outflows with the minor axis ($\Phi = 90^\circ$). 
To mimic observations, we create 2D images of galaxies in TNG50 using the stellar density. 
These images are fully idealised mocks that neglect observational effects such as noise, instrumental response and seeing and the effects of dust attenuation and scattering \citep{Bottrell2023}. 
We run \textsc{statmorph} \citep{Rodriguez-Gomez2019}, an algorithm utilised to determine the morphology of sources, on the mock images to model the galaxy profile and return the position angle (PA) and axis ratio, $b/a$. 
From here, we generate the azimuthal angles for each pixel in the galaxy projection, excluding the pixels within 5 kpc of the galaxy centre as $\Phi$ values are unreliable at small distances. 
For the figures in this section, we select only galaxies with $b/a$ values below the median of the sample. 
This is to exclude galaxies that may be at lower inclinations (more face-on) where the measured position angle is no longer reliable. 
We adopt this rather simple approach to measure the PA and $b/a$ because our goal is not to test the accuracy of azimuthal angle measurements in observations but rather, to provide a first-order approximation for what observers might measure. 

\begin{figure*}
    \includegraphics[width=1.0\textwidth]{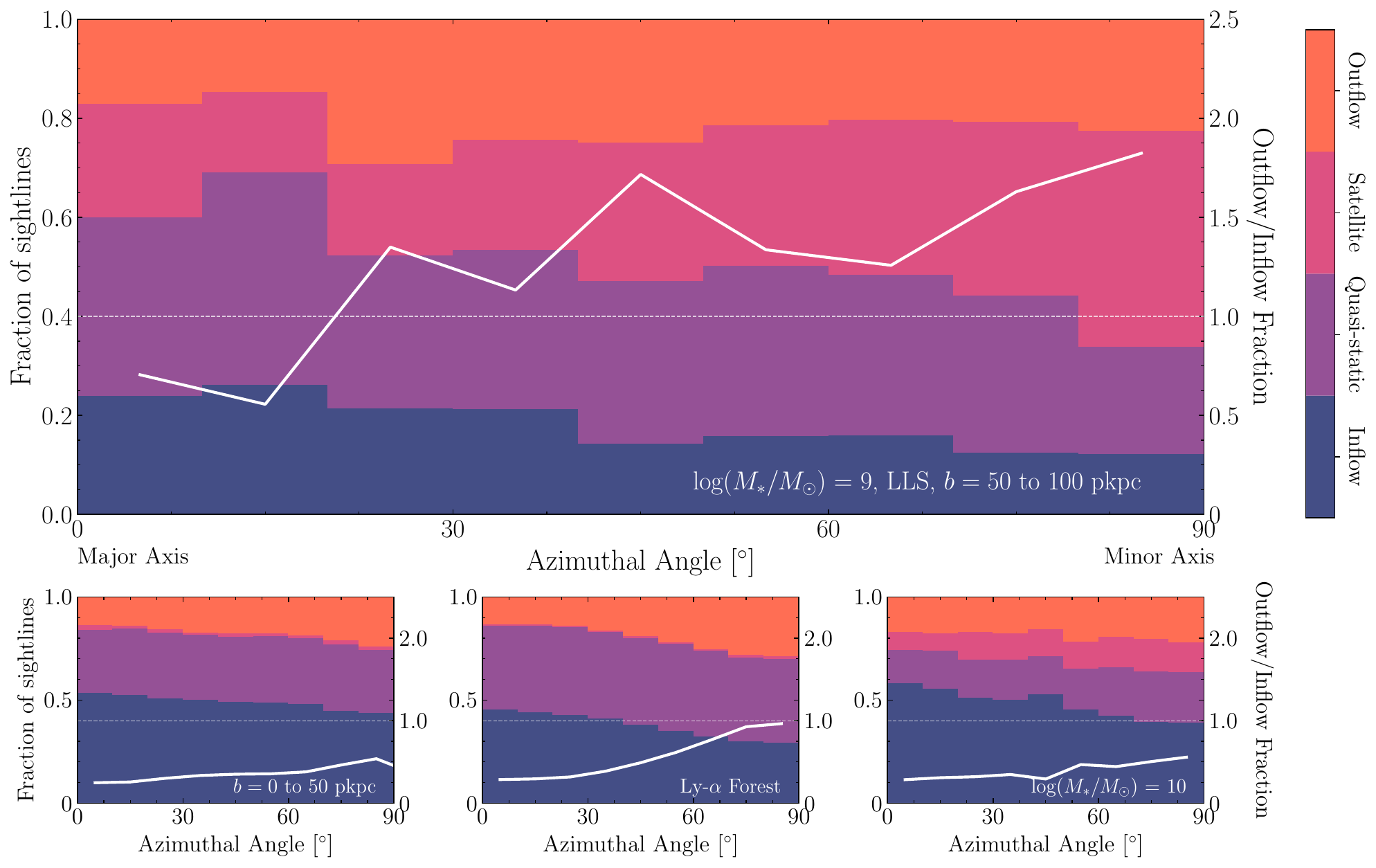}
    \caption{The fraction of absorbers intersecting gas originating from one of the four gas flow flags as a function of the azimuthal angle. 
    We use a consistent radial velocity cutoff of $\pm20$ \kms\ to label inflows and outflows and consider only central galaxies with a major-to-minor axis ratio less than the median value for the sample. 
    Using bins in $\Phi$ of 10$^\circ$, we colour each bin by the fraction of absorbers belonging to categories seen in the colour bar (left $y$-axis). 
    We find a significant contribution from gas that is in quasi-hydrostatic equilibrium (purple) at all azimuthal angles. 
    The solid white line represents the fraction of outflowing to inflowing absorbers (right $y$-axis) and the horizontal dashed line depicts an equal ratio. 
    There is an increase in this ratio as we move from the major axis to the minor axis. 
    This is most prominent in the primary panel where we consider LLS absorbers found at an impact parameter of 50 to 100 kpc from central galaxies of stellar mass $\sim$10$^9$ M$_\odot$.
    However, accounting only for absorbers at 0 to 50 kpc (bottom left) or absorbers around $\sim$10$^{10}$ M$_\odot$ central galaxies (bottom right) produces a minor difference of $\approx$0.2 in the outflow to inflow ratio between the major and minor axes. 
    While there is a clear increase in the outflow to inflow ratio towards larger $\Phi$ at lower \ion{H}{i} column densities (bottom middle), this ratio never exceeds unity. 
    }
    \label{fig:Sec5FracyesBAcut}
\end{figure*}

We show the fraction of absorbers that belong to one of the four gas flow flags as a function of the azimuthal angle in \autoref{fig:Sec5FracyesBAcut}. 
The shaded background conveys the fraction of absorbers belonging to the four categories using bins of ten degrees in azimuthal angle. 
The solid white line depicts the fraction of outflows to inflows, also as a function of the azimuthal angle where the horizontal dashed line marks an equal ratio. 
In the primary panel, we consider Lyman-limit systems found at an impact parameter of 50 to 100 kpc from 10$^9$ M$_\odot$ central galaxies. 
This cut in column density enables us to maximise the signal as higher $N_{\ion{H}{i}}$ systems are much rarer tracers of gas flows. 
In the bottom panels, we consider smaller impact parameters (left), lower column densities (middle) and larger stellar masses (right). 

We see a clear trend in the primary plot where the outflow to inflow ratio increases towards larger azimuthal angles. 
While similar signals have been detected before in simulations \citep{Nelson2019a, Mitchell2020}, previous studies calculate the azimuthal angle of pixels in a frame where the galaxy is viewed directly edge on. 
The method used in this paper, based on observational practices, only loosely constrains the inclination of the galaxy by considering axial ratios that are below the median value. 
This is a reassuring confirmation of observational practices where one uses $\Phi$ to distinguish gas flows \citep{Schroetter2016, Zabl2019}. 
Additionally, we also account here for the contribution from satellite galaxies and gas in quasi-hydrostatic equilibrium. 
Still, we confirm that the ratio of outflows to inflows is $\approx$0.5 near the major axis, rising to a peak of $\approx$2 at the minor axis in the primary panel. 

In the main panel, we have considered a particular subset of absorbers (LLS at impact parameters 50 to 100 kpc) associated with $M_* = 10^{9}$ M$_\odot$ central galaxies. 
When varying the impact parameter, \ion{H}{i} column density and central stellar mass, we obtain more varied results. 
Looking at the bottom left panel of \autoref{fig:Sec5FracyesBAcut} where we consider only absorbers within the inner 50 kpc of the galaxy, we still find an increasing outflow to inflow ratio, but the proportion of inflows dominates at all azimuthal angles. 
A similar signal is found when considering typical Ly-$\alpha$ forest column densities ($13.0 < \logNHIunit < 16.0$, bottom middle) and stellar mass (bottom right), namely, inflows dominate in number even near the minor axis. 
The influence of impact parameter and stellar mass are particularly significant, with only half the gas flows at the minor axis attributed to outflows. 
In particular, we find that accretion dominates sightlines passing through central galaxies belonging to the $M_* \approx 10^{10}$ and 10$^{11}$ M$_\odot$ bins. 
Current observational studies, which are more sensitive to massive galaxies, find signatures of outflows more common than inflows \citep{Martin2012, Rubin2012, Peroux2017, Rahmani2018a, Zabl2019, Weng2023b} and we discuss the possible reasons for this in Section \ref{sec:7discuss}. 

Looking specifically at the fraction of gas in quasi-hydrostatic equilibrium, we find little evolution with azimuthal angle. 
The bottom left panel of \autoref{fig:Sec5FracyesBAcut} shows that the fraction does not vary at lower impact parameters and only mildly increases for lower column densities (\textcolor{blue}{bottom middle}). 
Due to an increasing fraction of inflowing absorbers, there is much less gas in quasi-hydrostatic equilibrium for $M_* = 10^{10}$ M$_\odot$ central galaxies (bottom right). 
Curiously, we find a tentative signal of an increasing satellite LLS fraction at $50 < b < 100$ kpc as we move towards the minor axis for both $M_* = 10^9$ and $10^{10}$ M$_\odot$ central galaxies. 
For these stellar masses where outflows are typically starburst-driven, the gas density is expected to be greater along the minor axis. 
Hence, there should be stronger ram pressure stripping, leading to more quiescent and less gas-rich satellites near $\Phi = 90^\circ$ in the CGM. 
We find instead that satellites along the minor axis are more gas-rich, reminiscent of the anisotropic galaxy quenching found in galaxy clusters \citep{Navarro2021, Makoto2023}. 
A possible cause for this is if satellite galaxies along the major axis have been accreted at earlier times and hence, more likely to be quenched \citep{Karp2023}. 
At this stage, the origins of this anistropic LLS fraction from satellites remain ambiguous.  

\begin{figure*}
    \includegraphics[width=1.0\textwidth]{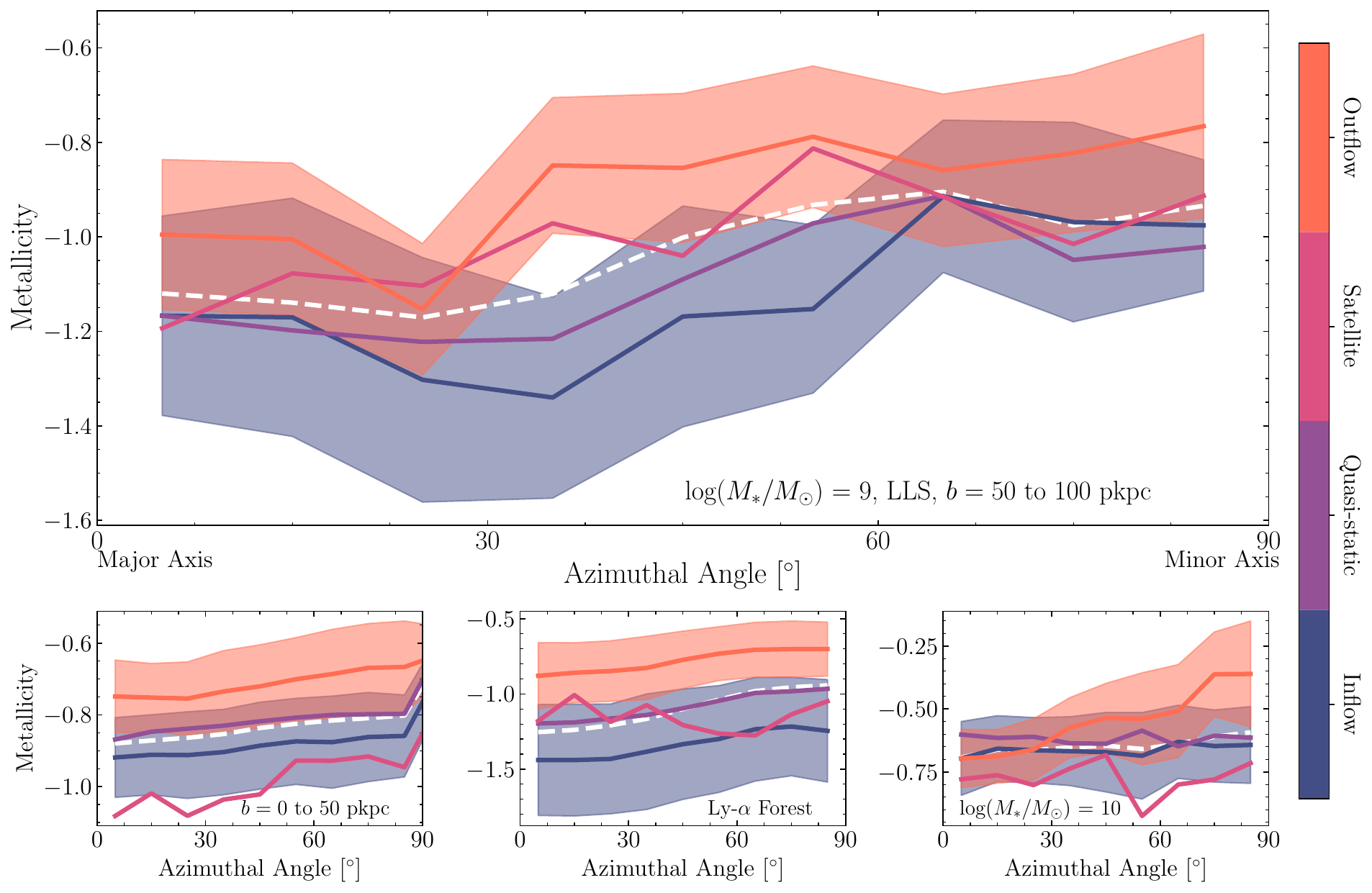}
    \caption{The metallicity with respect to solar of absorbers belonging to the four gas flow categories as a function of the azimuthal angle are plotted as solid lines. 
    The median metallicity of all absorbers is plotted in white. 
    We use a radial velocity cutoff of $\pm20$ \kms\ and only consider central galaxies with a major-to-minor axis ratio less than the median value for the sample.
    The shaded regions reflects the $0.5 \sigma$ uncertainty in the metallicity for only the inflowing and outflowing absorbers. 
    Absorbers associated with quasi-static gas and satellites do not have their errors plotted for clarity. 
    In the primary panel, we adopt the same impact parameter (50 to 100 kpc), column density (LLS) and stellar mass bins ($M_{\rm central} \approx 10^9$ M$_\odot$) as in \autoref{fig:Sec5MetyesBAcut}. 
    For the secondary panels, we also make the same variations from the fiducial parameters. 
    The positive gradient of the white dashed line in all panels shows that the median metallicity of absorbers increases towards the minor axis regardless of physical origin, although the effect diminishes for more massive central galaxies. 
    When we separate the absorbers into the four gas flow flags and plot their metallicities (coloured lines), we discover that the inflowing and quasi-static gas also increase in $Z$ at larger $\Phi$. 
    These absorbers then drive the trend in median metallicity as they are typically more abundant than outflowing absorbers.  
    }
    \label{fig:Sec5MetyesBAcut}
\end{figure*}

In \autoref{fig:Sec5FracyesBAcut}, we also find that 40 to 60 per cent of gas at all azimuthal angles arises from gas that is in quasi-hydrostatic equilibrium or in satellites. 
The contribution from satellites is most prominent in the primary panel which is consistent with our findings in \autoref{fig:Sec4SATfcover} where satellites dominate around $0.5$-$1 R_{\rm vir}$. 
At smaller impact parameters and lower column densities, their contribution diminishes. 
The contribution from quasi-static gas is significant for impact parameters up to 200 kpc, all central galaxy stellar masses and \ion{H}{i} column densities tested in this study. 
While the ratio of outflowing to inflowing sightlines increases with larger azimuthal angles, this value ignores the contribution from gas moving at slower radial speeds. 
{These results show that a fraction of absorbers in galaxy haloes may be associated with gas that is static or rotating. 
The latter has been observed in 60 per cent of Ly-$\alpha$ absorbers at $z \lesssim 0.03$ \citep{French2020} and recent simulations show that rotational support is significant in the CGM \citep{Lochhaas2023}. 
In this work, we have set an arbitrary restriction on the radial velocity ($\pm 20$ \kms) to distinguish between inflows, quasi-static gas and outflows. 
Hence, the fraction of gas that is rotating without loss of angular momentum (quasi-static) compared to co-rotating inflows depends on the radial velocity boundary.}
Any increase in magnitude of the $v_r$ cutoff used naturally leads to an increase in the fraction of gas in hydro-static equilibrium. 
Likewise, setting all negative (positive) radial velocities to be counted as inflows (outflows) leads to no absorbers being labelled as quasi-static. 
However, we find that unless the $v_r$ cutoff is $> 150$ \kms, the ratio of inflows to outflows remains similar; there is just a proportional increase in the quasi-static absorber fraction. 
At high radial velocities, we only expect to find outflows and hence, the fraction of inflows to outflows decreases dramatically. 
Therefore, we caution that interpreting \ion{H}{i} absorbers as inflowing or outflowing using their projected azimuthal angle relative to the major axis should include further consideration of other properties such as $b$, $\logNHI$ and $M_*$. 

\subsection{The metallicity of gas flows}
Another absorber property used to differentiate outflows from inflows is the gas phase metallicity \citep{Fox2016, Fox2019, Ramesh2023b}. 
In \autoref{fig:Sec5MetyesBAcut}, we plot the metallicity as a function of the azimuthal angle. 
The four coloured lines correspond to the four gas flow origin flags and the dashed white line is the median metallicity of all absorbers in a given azimuthal angle bin with size 10$^\circ$. 
The shaded regions indicate the 0.5$\sigma$ uncertainty in the metallicity for only outflows and inflows. 
We adopt the same fiducial parameters (LLS found 50 to 100 kpc from $M_* = 10^9$ M$_\odot$ central galaxies) in the primary panel as \autoref{fig:Sec5FracyesBAcut} and the same changes in the secondary panels. 

For almost all azimuthal angles and choices of $b$, $N_{\ion{H}{i}}$ and central galaxy stellar mass, the metallicity (with respect to solar) of outflows is consistently 0.2 to 0.5 dex larger than inflows. 
The few exceptions occur for absorbers at azimuthal angles $\Phi < 30^\circ$ found towards $M_* = 10^{10}$ M$_\odot$ central galaxies (bottom right) where the accreting gas is roughly equal in metallicity to the outflowing gas, perhaps due to recycling. 
We also find that absorbers are more metal-rich around more massive central galaxies, while the impact parameter and \ion{H}{i} column density change the normalisation by $< 0.1$ dex. 
It is for this same reason that the metallicity of satellites is typically lower than the median value; as satellites are usually less massive, their average metallicities are also going to be lower. 

Similar to the findings of \citet{Peroux2020, vandevoort2021}, we find a positive gradient in the median metallicity of absorbers as a function of azimuthal angle. 
The magnitude of the difference in $Z$ between the minor and major axes is typically of order $0.2$ dex, but does diminish with increasing central stellar mass. 
More curiously, we find that the driver of this metallicity difference is only partly caused by an increase in the fraction of absorbers tracing metal-enriched outflows. 
In general, gas that is inflowing or in quasi-hydrostatic equilibrium also increases in metallicity towards the minor axis, hinting that the pollution of metals into the circumgalactic medium near the minor axis also results in metal-enriched accretion in the form of recycled gas \citep{Fraternali2017, Muratov2017}. 
As the fraction of quasi-static and inflowing gas typically dominates the outflow fraction (bottom panels in \autoref{fig:Sec5FracyesBAcut}), the increase in median metallicity of these types of absorbers contribute more significantly to the median metallicity of the CGM. 
We discuss the metallicity distribution of inflowing and outflowing gas in more detail in Section \ref{sec6:obs}.  

\section{Observational Comparisons}
\label{sec6:obs}
\subsection{Determining the host galaxy of absorbers}
With the proliferation of IFS surveys targeting absorbers, there are an increasing number of cases where multiple galaxies are found within some velocity cutoff with respect to the absorber \citep[e.g.][]{Dutta2020, Hamanowicz2020, Berg2023, Weng2023a}. 
This points to a more complex relationship between galaxies and absorbers, and one that may only be captured by larger statistical studies. 
At the same time, it is also critical to relate the properties of the CGM with galaxy properties in individual cases to understand how galaxies evolve \citep[e.g.][]{Tumlinson2013, Burchett2016, Werk2016, Langan2023}. 
However, the fidelity of our current methods used to associate galaxies with absorbers remains ambiguous. 
We have shown already in Figures \ref{fig:Sec3dLOS} to \ref{fig:Sec4LowMassSat} the importance of considering other haloes along the line of sight and $M_* < 10^8$ M$_\odot$ satellites. 
The predominant method of assuming the galaxy at lowest impact parameter to the QSO sightline is the host neglects both these considerations. 
Here, we analyse the potential host galaxy of absorbers using the results of these simulations and prescribe a more reliable method for associating absorbers to galaxies. 


\begin{figure*}
    \includegraphics[width=1.0\textwidth]{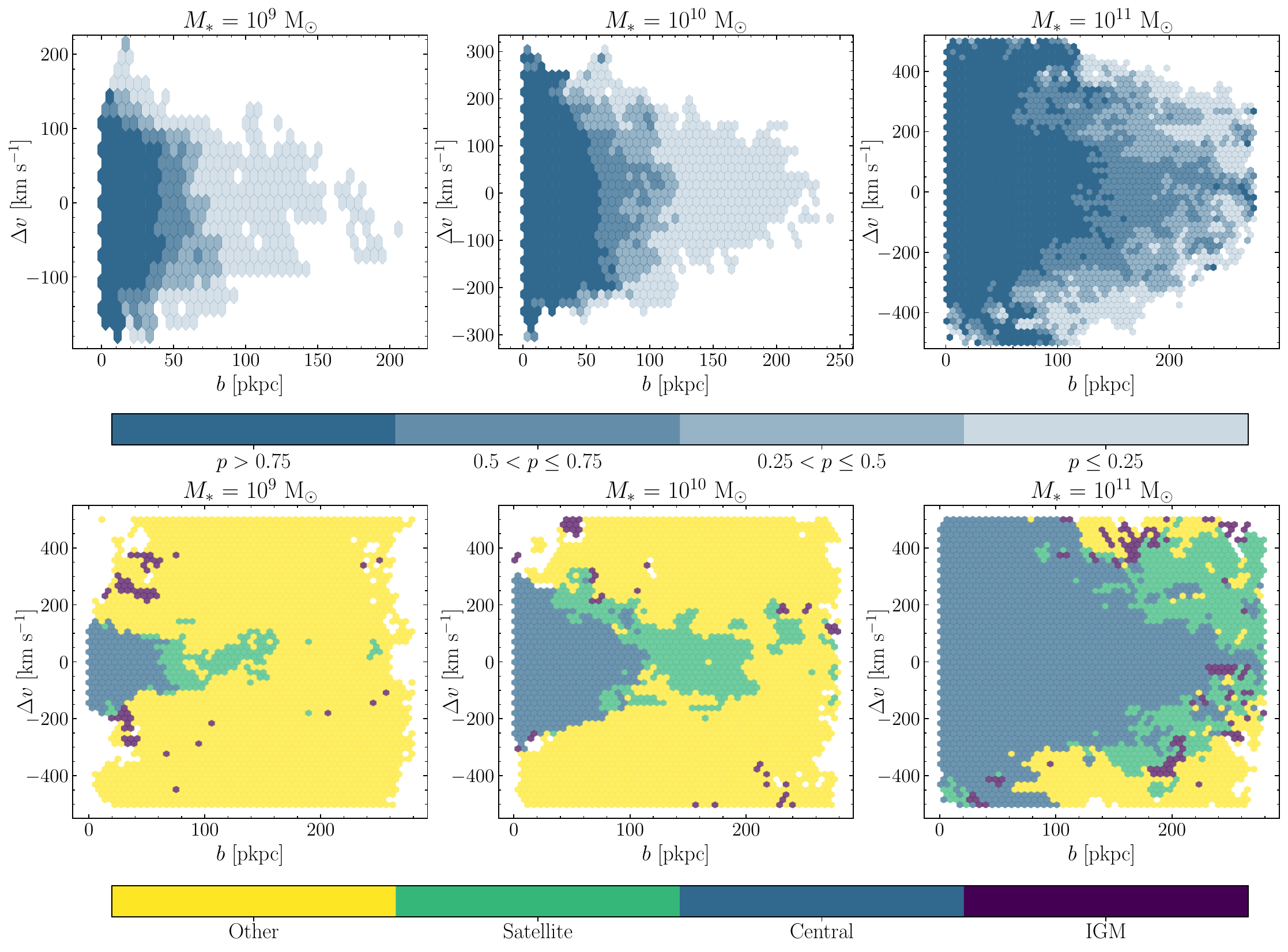}
    \caption{{In the top row, we show the fraction of $\logNHIunit > 18.0$ absorbers belonging to the central galaxy given its impact parameter ($x$-axis) and line of sight velocity difference ($y$-axis). 
    In the bottom row, we show the regions where satellites, other haloes and the IGM dominate (note the differing axis ranges).  
    Hexbins are shaded by the colour corresponding to the most likely origin in the bottom row; we emphasise that the origin is probabilistic which we demonstrate in the top row. 
    We use four likelihood bins [$p > 0.75$, $0.5 < p \leq 0.75$, $0.25 < p \leq 0.5$, $0 < p \leq 0.25$] to colour each hexbin by the corresponding probability ($p$) that the central galaxy hosts the absorber at some impact parameter and velocity. 
    There are three plots for the stellar mass bins: $10^9$ (left) $10^{10}$ (centre) and $10^{11}$ (right) M$_\odot$. 
    It is clear that the central galaxy's stellar mass drives the coverage of \ion{H}{i} in $\Delta v$-$b$ space. 
    This figure represents a step forward from assuming the galaxy at lowest impact parameter is the host of the absorber by taking into account both $b$ and $\Delta v$ and can be used to guide observers when selecting counterparts to absorbers. 
    At the same time, we emphasise that the contamination from origins such as satellites and other haloes are significant. 
    For current and future surveys of \ion{H}{i} absorbers, we consider different \logNHI bins in \autoref{fig:AppProb}.}
    }
    \label{fig:Sec6MAHadvise}
\end{figure*}

In \autoref{fig:Sec6MAHadvise}, we show the most likely origin flag of an absorber found at some impact parameter and line of sight velocity difference to the central galaxy. 
The likelihood changes significantly as a function of stellar mass and the three panels represent stellar masses $M_* = [10^9, 10^{10}, 10^{11}] \rm{M}_\odot$ from left to right. 
We also impose a \ion{H}{i} column density limit of $\logNHI > 18.0$; both the stellar mass and column density restrictions are intended to match the MUSE-ALMA Haloes sample (black stars). 
At lower column densities, there is an increasing contribution from the intergalactic medium, replacing the contribution from other haloes and satellites. 
However, the extent of the central galaxy (blue) remains similar as a function of $\Delta v$ and $b$, even when considering all absorbers with $\logNHIunit > 13.0$. 
If we look at the distribution of absorbers around galaxies from the COS-Halos survey \citep{Tumlinson2013, Werk2013, Werk2014, Werk2016, Peeples2014}, we find that the centroids of the total \ion{H}{i} absorption profile are predominantly within the region dominated by the central galaxy after matching stellar masses. 
However, the centroids of individual \ion{H}{i} components are occasionally found at large velocities from the galaxy systemic redshift and are more likely associated with other haloes or the IGM (if at lower column density). 
The line of sight velocity difference between absorber and galaxy remains an important consideration and we suggest that when associating galaxies with absorbers, both $\Delta v$ and $b$ are considered \citep[e.g.][]{Berg2023}. 

While we have coloured each hexbin by a given flag, we emphasise that this is inherently probabilistic and there is much uncertainty, particularly at the boundaries between the central and satellites or other haloes.  
Though \autoref{fig:Sec6MAHadvise} is a useful observational diagnostic for determining the potential origins of \ion{H}{i} absorbers by moving beyond just using the impact parameter, it underscores that absorbers may not have a single well-defined origin and the need for statistical studies involving larger samples. 
This point is implicitly made by works studying absorbers arising from the intragroup medium where absorbers are not associated with a single galaxy \citep{Gauthier2013, Nielsen2018, Klitsch2018, Peroux2019, Dutta2021, Dutta2023, Qu2023}. 

Thus far, we have adopted a galaxy-centric approach, that is, we study how the properties of \ion{H}{i} absorbers vary as we move from a selected central galaxy. 
Many of the current IFS absorber studies are centred on a background source which is arbitrarily positioned in the sky with respect to foreground galaxies \citep[e.g.][]{Schroetter2016, Chen2020, Hamanowicz2020, Lofthouse2020, Muzahid2020, Nielsen2020, Muzahid2021, Berg2023}, including the MUSE-ALMA Haloes survey \citep{Peroux2022}. 
To produce results that can be directly used to better understand observations of this type, we select a random pixel within a 200 kpc square centred on the central that has a column density $\logNHIunit > 18.0$. 
This becomes the location of our absorber and we calculate the impact parameter of the remaining pixels in the field with respect to the absorber. 
We can then determine the fraction of absorbers belonging to the four origin flags in a similar manner to \autoref{fig:Sec4SATfcover}. 
For each mock, we repeat this process five times for each projection direction, leading to 15 iterations in total for each central subhalo. 

\begin{figure*}
    \includegraphics[width=1.0\textwidth]{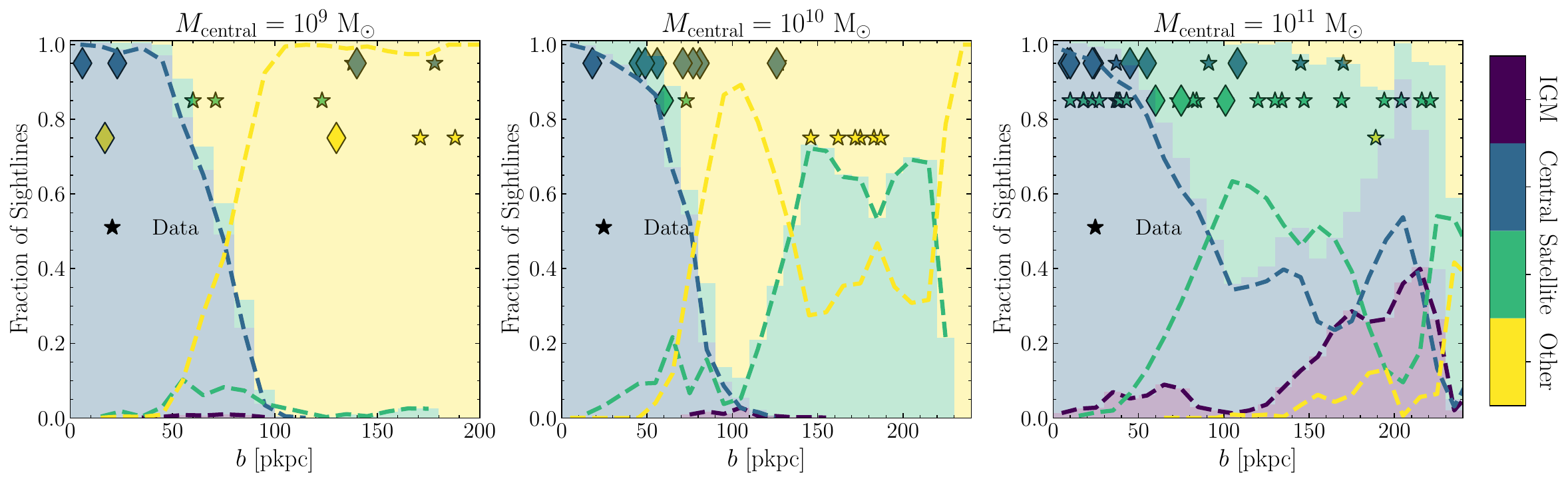}
    \caption{The fractional pixel origins ($y$-axis) given its impact parameter ($x$-axis) from a randomly-selected $\logNHIunit > 18.0$ absorber. 
    There are three plots for the stellar mass bins: $10^9$ (left) $10^{10}$ (centre) and $10^{11}$ (right) M$_\odot$. 
    The dashed lines show how the fraction of pixels tracing the central, satellites, other haloes and the IGM vary as a function of impact parameter. 
    Using bins of 10 kpc in $b$, we also highlight their relative fractions using the coloured background. 
    Data points from the MUSE-ALMA Haloes survey are plotted as stars and diamonds (for the galaxy nearest the absorber) using the publicly available catalogue from \citet{Weng2023a}. 
    We find a rough agreement between the abundances of central, satellite and other halo galaxies and the predictions from TNG50. 
    The excess of satellite galaxies at low impact parameters for the $10^{11}$ M$_\odot$ (right) plot arises from a large galaxy group consisting of $> 10$ galaxies found towards Q1130$-$1449 \citep{Peroux2019, Chen2019}.
    }
    \label{fig:Sec6MAHFracCompare}
\end{figure*}

The results are shown in \autoref{fig:Sec6MAHFracCompare}. 
Using a randomly-selected absorber, we plot how the fraction of pixels tracing the central, satellites, other haloes and the IGM surrounding the absorber changes as a function of impact parameter. 
The dashed lines corresponding to the four origins show this variation with $b$ and the coloured background elucidates the relative fractions within each 10 kpc impact parameter bin. 
To make an appropriate comparison with the MUSE-ALMA Haloes survey, we re-analyzed this observational data and mimicked the definitions of centrals, satellites and other haloes used in TNG50. 
We derive the halo mass and virial radius using the stellar-to-halo mass relation \citep{Puebla2017}.  
We define the central galaxy as the most massive galaxy in the MUSE field of view and hence, other galaxies associated with the absorber are described as satellites (other haloes) if they are inside (outside) $R_{\rm vir}$ of the central \citep{Karki2023}. 
While satellites can still be bound beyond $R_{\rm vir}$, we choose to adopt this simpler definition for this exercise. 
Then, we apply a column density cutoff of $\logNHIunit > 18.0$ and only consider $M_* = 10^9$ to $10^{11}$ M$_\odot$ central galaxies to match the MUSE-ALMA Haloes survey. 
The impact parameter of these various galaxies in the data are plotted using stars with a filled colour corresponding to their origin. 
For clarity, we plot central galaxies in the top row, followed by satellites and other haloes as we move down. 

The relative abundances of central, satellite and other halo galaxies as a function of impact parameter from the MUSE-ALMA Haloes survey is broadly consistent with the predictions from TNG50. 
This highlights that the most massive halo is not necessarily the host galaxy of the absorber and that satellites and other haloes can be significant contributors to the \ion{H}{i} gas down any line of sight. 
A larger sample size, particularly for the $10^9$ M$_\odot$ central galaxies, is required to make more quantitative comparisons. 
We note that the excess of satellites at low impact parameters for the $10^{11}$ M$_\odot$ central galaxies is caused by the DLA towards Q1130$-$1449 being associated with more than 10 galaxies \citep{Peroux2019, Chen2019}. 
The right plot of \autoref{fig:Sec6MAHFracCompare} also highlights that gas ejected from AGN that is no longer gravitationally bound may contribute significantly to the covering fraction around massive haloes. 
While this signal may be enhanced by the kinetic feedback mode of AGN in the IllustrisTNG model pushing large quantities of cool gas into the CGM, it highlights the potential for absorbers to be unbound to any galaxy which is ignored in most observational studies \citep[see][for an exception]{Berg2023}.

\subsection{Metallicity distribution of \ion{H}{i} absorbers}
We now move to statistical studies on absorber metallicities and their possible origins in and around galaxies. 
The distribution of pLLS and LLS absorber metallicities at $z \lesssim 1$ appears to be bimodal from observations \citep{Lehner2013, Lehner2019, Wotta2016, Wotta2019, Berg2023}. 
In \citet{Lehner2013}, the authors find tentative evidence for a metallicity bimodality using absorbers with column densities $16.2 \lesssim \logNHIunit \lesssim 18.5$ at $z \lesssim 1$. 
In a larger sample, \citet{Wotta2016, Wotta2019} find that the bimodality is driven by pLLS absorbers where the metallicity peaks at [X/H] $=-1.7$ and $-0.4$. 
This bimodality is purported to arise from low-metallicity absorbers tracing gas accretion or overdense regions of the Universe \citep{Berg2023}, while more metal-rich absorbers are associated with the CGM of galaxies. 
{However, a study of $L^*$ galaxies from the COS-Halos Survey finds a unimodal metallicity distribution centred on [X/H] $\approx-0.5$ \citep{Prochaska2017} and argues that low-metallicity absorbers are not necessarily freshly accreted gas.} 
As our TNG50 sightlines are centred on central galaxies with specific halo masses, we do not attempt to reproduce the number counts in these observations \citep[see][]{Hafen2017, Rahmati2018}, but rather investigate whether a bimodality exists in the normalized sense using the various absorber origins. 

\begin{figure*}
    \includegraphics[width=1.0\textwidth]{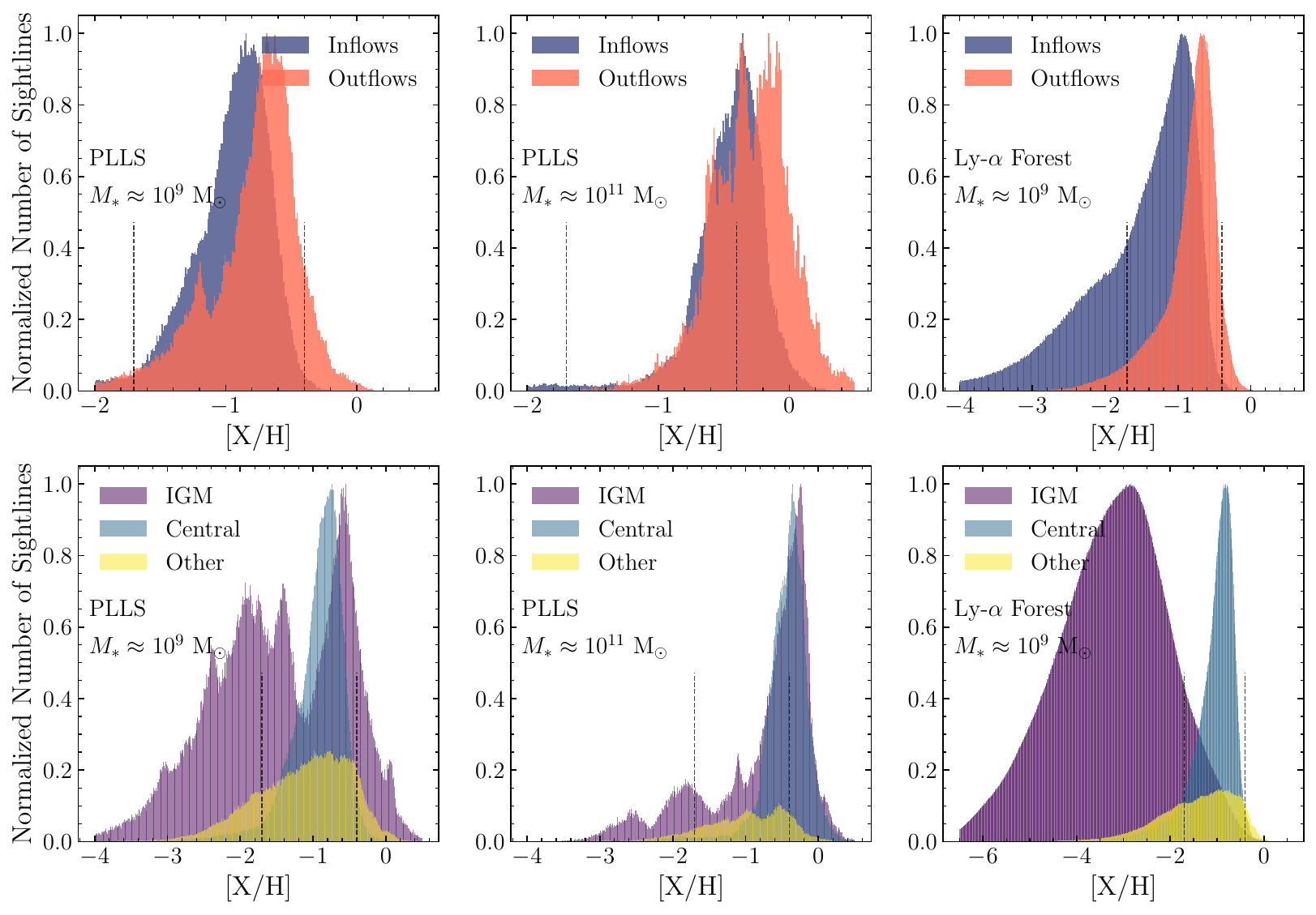}
    \caption{The normalized distribution of absorber gas phase metallicities. 
    We adopt \ion{H}{i} column densities $16.0 < \logNHIunit < 19.0$ to match observational samples and a central stellar mass of $M_* = 10^9$ M$_\odot$ as the fiducial parameters (left panels). 
    The vertical dashed lines indicate the peaks of the bimodal metallicity distribution found in observations. 
    In the top (bottom) panels, we separate sightlines probing inflows (the central) from outflows (the IGM and other haloes) and display the distribution of metallicities for the separate populations. 
    The histograms have been normalized such that the peaks of each distribution are at unity. 
    {An exception is for `other halo' galaxies in the bottom row which have been normalized by the peak of the central galaxy metallicity distribution for clarity. }
    There is a $\approx$0.2 dex difference in metallicity between inflows and outflows for the fiducial parameters. 
    At higher stellar masses (lower column densities), this difference diminishes (increases) in magnitude. 
    We find gas in the IGM spans a broad range of metallicities, from metal-poor to metal-rich, due to significant environmental enrichment from nearby galaxies ejecting metals beyond their haloes. 
    {Similarly, gas belonging to other haloes along the line of sight will be of varying stellar masses and hence, have a significant range of metallicites.}
    }
    \label{fig:Sec6MetDist}
\end{figure*}

In \autoref{fig:Sec6MetDist}, we display the metallicity distributions of absorbers separated by their origin in TNG50. 
We first separate absorbers by whether they trace inflows or outflows in the top panels and then, absorbers tracing the IGM from absorbers tracing the central galaxy in the bottom panels. 
All distributions are normalized to have equal peak heights and do not reflect the actual incidence of absorbers. 
While outflowing gas is typically more metal-rich than inflowing gas, we find only a $\sim$0.2 dex different in their metallicity distribution peaks for the fiducial parameters of $16.0 < \logNHIunit < 19.0$ and $M_{\rm cen} = 10^9$ M$_\odot$. 
This difference increases to $\sim$0.4 dex if we only consider $13.0 < \logNHIunit < 16.0$ absorbers (top right), but almost disappears when considering $M_{*} = 10^{11}$ central galaxies (top centre). 
Regardless, any difference we find is minor when compared to observations where the difference in $[\rm{X/H}]$ is 1.3 dex. 
The strong overlap in metallicities is in part caused by efficient wind recycling, where previous metal-enriched outflows later become inflows \citep{Daniel2017}. 
We choose to not include absorbers that are in quasi-hydrostatic equilibrium or associated with satellites in this sample for clarity but note that their presence further dilutes any signal of bimodality (as seen in the unimodal `central' distributions in the bottom panels). 
These results are consistent with previous simulated metallicity distributions \citep{Hafen2017, Rahmati2018} and in tension with some observations \citep{Fumagalli2016, Wotta2019}. 

In the bottom panels of \autoref{fig:Sec6MetDist}, we compare the metallicity distributions of gas tracing the intergalactic medium with gas bound to the central galaxy. 
Curiously, we find absorbers with $16.0 < \logNHIunit < 19.0$ and metallicities $[\rm{X/H}] > 0.1$ that trace gas in the IGM. 
When we compare these values with the gas bound to a $M_* = 10^9$ M$_\odot$ central galaxy (bottom left), we find the IGM gas is metal-enriched with respect to the galaxy. 
These metal-enriched IGM absorbers arise from the expelled material of more massive haloes found along the sightline, as feedback processes drive metal enriched gas to large distances from the central regions of galaxies \citep[e.g.][]{Madau2001,Scannapieco2002, Cen2011, Maio2011, Ayromlou2023}. 
This is why we see an alignment in metallicity distributions between the peak IGM component and the $M_* = 10^{11}$ M$_\odot$ central galaxy sample in the bottom middle panel. 
When we select for \ion{H}{i} column densities $\logNHIunit < 16.0$, we see a 3 dex discrepancy between the distribution peaks, with IGM absorbers found at lower $Z$ as expected. 

{Whether the metallicity distribution of partial Lyman-limit systems is unimodal or bimodal at $z = 0.5$ is tied directly to the incidence of absorbers belonging to the various origins (e.g. IGM and central). 
We reiterate that any direct comparison with observational studies is difficult as our sightlines are inherently found near galaxies. 
However, we do find that in the TNG50 simulation, the metallicity is not a perfect discriminator of gas that is accreting or outflowing \citep[consistent with the findings of][]{Hafen2017, Rahmati2018}. 
Moreover, the peak of the metallicity distribution for absorbers associated with $10^{10}$ and $10^{11}$ M$_\odot$ central galaxies is consistent with both the metal-rich peak from the bimodal distribution of \citet{Wotta2016} ([X/H] $\approx -0.4$) and the peak of the unimodal distribution ([X/H] $\approx -0.5$) in \citet{Prochaska2017}. 
While the peaks align, \citet{Prochaska2017} finds a spread in metallicities around $L^*$ galaxies that is roughly twice as large as predicted in TNG50. 
In \autoref{fig:Sec6MetDist}, we observe that the increased spread towards lower [X/H] may arise from less-massive satellite or secondary halo galaxies that are typically lower metallicity from the mass-metallicity relation \citep{Lequeux1979}. 
We note that satellite galaxies have not been shown for clarity, but their metallicity distribution also extends to smaller [X/H] values. 
The lower incidence of super-solar metallicities ([X/H] $> 0.5$) may be caused by the limited CGM resolution in TNG50 ($\sim$ 1 pkpc beyond $0.1R_{\rm vir}$), although we note that \citet{Wotta2016, Wotta2019} find an upper limit in [X/H] of roughly 0.5. 
Ultimately, our results suggest that inferring the origin of gas from metallicity requires caution. 
For random sightlines towards UV-bright QSOs \citep[e.g.][]{Lehner2018}, we find that gas expelled from massive haloes into the intergalactic medium are a source of metal-enriched absorbers. 
Likewise, the large spread in [X/H] values around $L^*$ galaxies may imply that absorbers are perhaps associated with lower-mass satellites that have lower metallicity. }

\section{Discussion}
\label{sec:7discuss}
{\subsection{The CGM of star-forming and quiescent galaxies}
The analysis in this work focuses on the CGM properties of galaxies with varying stellar masses at $z = 0.5$, coinciding with the $M_*$ range and redshift of galaxies in MUSE-ALMA Haloes. 
At any given stellar mass bin, there are galaxies above, below and on the SFR-$M_*$ main sequence. 
Here, we discuss how the CGM properties of galaxies with similar stellar mass depend on the star-formation rate.} 

{
The central galaxies in this work are randomly selected from TNG50 and enforced to have stellar masses within 0.3 dex of $M_* = [10^{8.0}, 10^{9.0}, 10^{10.0}, 10^{11.0}]$\,$\rm{M_\odot}$. 
In total, there are  [100, 100, 50, 20] galaxies within each bin, and we further categorise the galaxies as star-forming or quiescent. 
The distinction is made somewhat arbitrarily; we consider galaxies with SFRs in the lowest quartile of each bin as quiescent, and the remainder are star-forming. 
We also tested separating star-forming and quiescent galaxies using the median SFR but found little variation in the results presented below. }

{
We find marginal differences in the results presented in this paper after separating star-forming and quiescent galaxies. 
The fraction of sightlines with \ion{H}{i} as a function of impact parameter for the various origins (\autoref{fig:Sec4SATfcover}) are within the uncertainties for both groups. 
Likewise, there are marginal differences in the fraction of inflows, quasi-static gas, outflows and satellites as a function of azimuthal angle between star-forming and quiescent galaxies (\autoref{fig:Sec5FracyesBAcut}). }

{
These results highlight that the stellar mass (because it is tied to halo mass) is the key driver of the amount, distribution and extent of the \ion{H}{i} in the circumgalactic medium. 
Additionally, it is unclear in observations whether the cool gas content in the CGM of star-forming and quiescent galaxies differ. 
From studies of luminous red galaxies, the cool gas mass in the CGM of passive galaxies appears comparable to star-forming galaxies \citep{Chen2018, Zahedy2019}. 
As \ion{H}{i} is almost ubiquitous in galaxy haloes \citep[e.g.][]{Tumlinson2013, Werk2013, Werk2014}, Ly-$\alpha$ is not the ideal absorption line to trace differences between the circumgalactic media of star-forming and passive galaxies. 
Instead, it appears that the \ion{O}{vi} covering fraction is significantly larger around star-forming galaxies \citep{Werk2013, Tchernyshyov2023}. 
To understand the different CGM properties of star-forming and quiescent galaxies in simulations, we require a more systematic study across all stellar masses rather than isolated bins in this work. 
The simplistic method of separating star-forming and passive galaxies used might weaken any potential differences in their CGM properties. 
However, we note that our results are consistent with observations and highlight that stellar (halo) mass drives the \ion{H}{i} amount, extent and distribution around galaxies. }

\subsection{The incidence of inflows in observations and simulations}
The incidence of inflowing gas observed in both tranverse absorption line and down-the-barrel observational studies is inferred to be $\approx$10 per cent \citep{Kacprzak2010, Martin2012, Rubin2012, Ho2017, Zabl2019, Weng2023b}. 
Comparatively, our TNG50 results indicate a minimum inflow fraction of $\approx$20 per cent which rapidly increases for higher stellar masses (\autoref{fig:Sec5FracyesBAcut}). 
We emphasise that observational studies of gas flows using absorption-lines towards background sources provide only fragmentary information about the physical origin of the observed gas and perhaps our assumptions on whether the gas traced is inflowing or outflowing require reassessment (see Section \ref{sec:flows}). 
Moreover, when we collapse the mock into a 2D projection, there is almost always both inflowing and outflowing gas in each sightline. 
This conflicts with assumptions made in most transverse absorption-line studies, where an entire absorption system is typically assigned to a single gas flow \citep[e.g.][]{Schroetter2019, Zabl2019, Weng2023b}.
In reality, the absorption could be produced by a combination of the two processes \citep[see][for an example]{Nielsen2022} and current observational methods do not capture this adequately.   
This may be one cause of the disparity in inflow incidences between observations and simulations. 
Despite these observational and modelling challenges, the incidence of sightlines tracing inflowing gas remains an important and potentially constraining quantity to measure. 


\subsection{The number of clouds along each sightline}
For the characterization of different origins, and measurements of absorber properties per sightline, we assume that each sightline is dominated by a single cloud of gas belonging to one of the flags. 
In reality, there is the chance of intersecting multiple clouds which we commonly find in absorption line observations. 
If there are multiple clouds along the line of sight found at different velocities, then the measurement of properties such as the line of sight distance or metallicity may be averaged out. 
However, \citet{Ramesh2023d} shows that the typical number of clouds intersected with $\logNHIunit > 16.0$ along a sightline is approximately two near the centre of Milky Way-like galaxies at $z = 0$, decreasing to a single cloud towards the virial radius for TNG50. 
Hence, at this resolution in the CGM, it is likely we are only intersecting a single cloud. 
In the event there are multiple clouds, we typically find that a single cloud dominates the \ion{H}{i} mass. 



\subsection{Extension to different gas phases}
\begin{figure*}
    \includegraphics[width=1.0\textwidth]{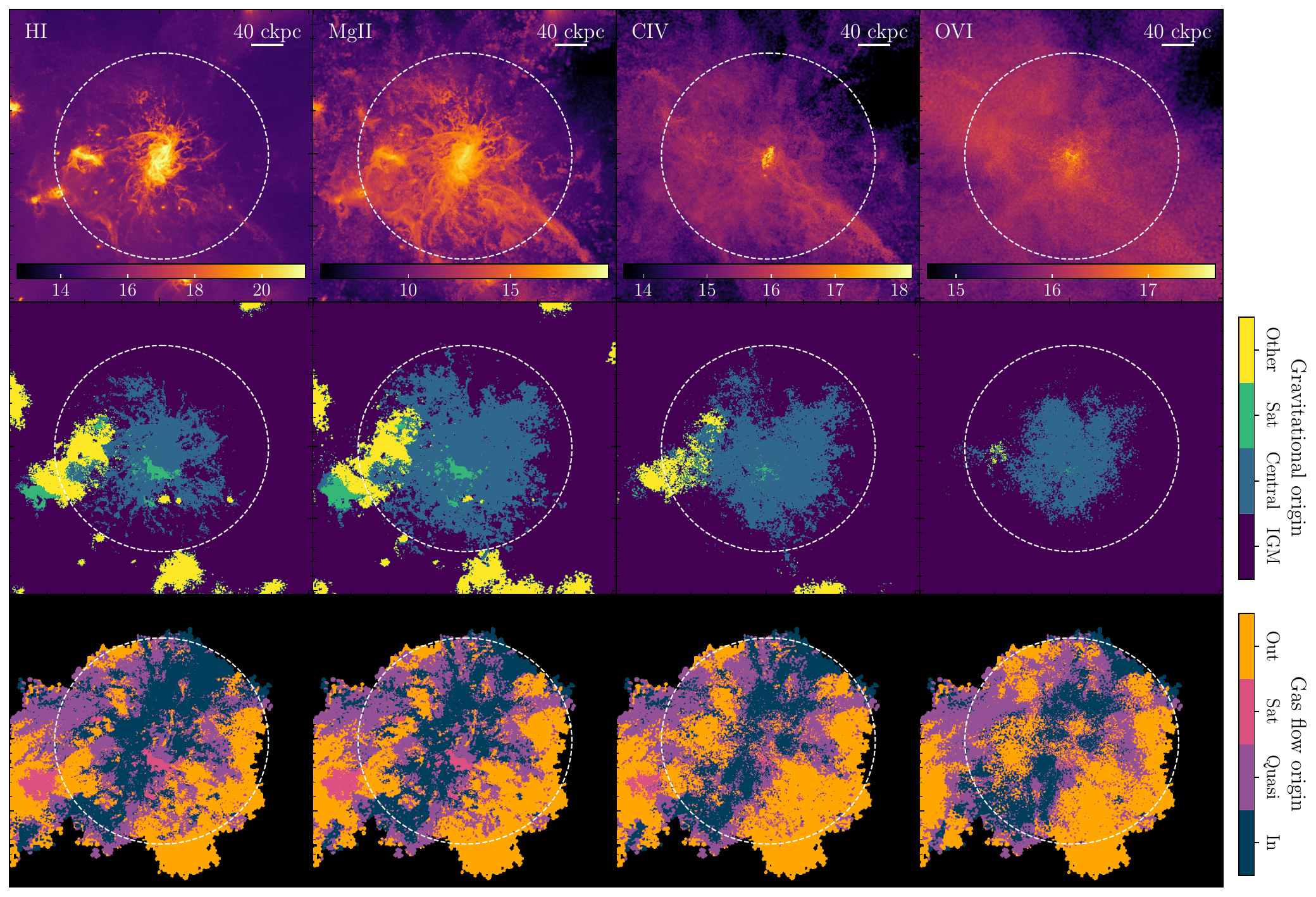}
    \caption{A visualisation of a $10^{10.2}$ M$_\odot$ central galaxy (subhalo ID 455488) in TNG50. 
    In the top row, we show the \ion{H}{i}, \ion{Mg}{ii}, \ion{C}{iv} and \ion{O}{vi} column density maps respectively from left to right. 
    The middle and bottom rows colour the pixels by the flag that dominates the \ion{H}{i} or ion mass in projection. 
    The dashed white circle outlines the virial radius of the galaxy. 
    We see that the extent, distribution and origin of absorbers depend significantly on the tracer used. 
    The \ion{C}{iv} and \ion{O}{vi} maps are dominated by the massive central galaxy, while we find more contributions from satellites and other haloes in the \ion{H}{i} and \ion{Mg}{ii} maps. 
    More of the gas is found to be outflowing in the higher-ionization tracers than in the \ion{H}{i} or \ion{Mg}{ii}, where the latter is particularly marginal in extent. 
    }
    \label{fig:Sec6OtherIons}
\end{figure*}

Thus far, we have analysed \ion{H}{i} absorbers in this study. 
Other common metal absorption lines used in studies of absorber counterparts include \ion{Mg}{ii}, \ion{C}{iv} and \ion{O}{vi}. 
In the top row of \autoref{fig:Sec6OtherIons}, we show column density maps of each ion for the same $M_* \approx 10^{10}$ M$_\odot$ galaxy displayed in \autoref{fig:Sec2TNGplot}. 
While the \ion{H}{i} and \ion{Mg}{ii} absorbers are densest at the centre of galaxies, the \ion{C}{iv} and \ion{O}{vi} absorbers that trace hotter gas phases are more diffuse and extended \citep{Ho2021, Nelson2023}. 
Following the previous analysis for \ion{H}{i}, we designate each pixel by the identical set of origin and gas flow flags using the aggregate \ion{Mg}{ii}, \ion{C}{iv} or \ion{O}{vi} mass along each sightline. 
Looking at the second row of \autoref{fig:Sec6OtherIons}, we find that the more massive central galaxy dominates most sightlines over satellite and other halo galaxies for \ion{C}{iv} and \ion{O}{vi}. 
This is caused by smaller haloes not reaching the virial temperatures ($\approx10^{5.5}$ K) required to form \ion{O}{vi}. 
The structure of outflows in the bottom row also differs between the ions. 
A visual inspection shows that ions tracing the hotter phases of gas have a larger proportion of sightlines dominated by outflows. 
This is in line with the earlier discussion of the high incidence of \ion{H}{i} inflows; because the outflowing gas is hotter, it is more likely to reach the temperatures required to produce \ion{C}{iv} and \ion{O}{vi}. 
We present here only some preliminary insights into how different absorber species have varying origins. 
A complete analysis will be released in an upcoming work. 

\section{Conclusions}
We analyse the physical origins of \ion{H}{i} absorbers in and around central galaxies with stellar masses 10$^8$ to 10$^{11}$ M$_\odot$ at $z = 0.5$ using the TNG50 simulation. 
We consider all gas cells within $\pm$ 500 \kms\ of the central galaxy and categorise all cells based on whether they are gravitationally bound to the central galaxy, a satellite of the central or another halo. 
Additionally, we also consider gas that is unbound and in the intergalactic medium. 
These four flags [IGM, central, satellite, other halo] form the origin labels. 
We also derive a second set of gas flow labels: [inflows, quasi-hydrostatic, satellites and outflows].
We then connect absorber properties such as impact parameter, line of sight velocity difference, line of sight distance, metallicity, azimuthal angle and column density with their origin or gas flow. 
Our major findings are summarised here. 
\begin{itemize}
    \item The line of sight velocity difference is a poor indicator of the physical distance between the absorber and galaxy, particularly if the absorber traces gas in the intergalactic medium or another halo near the sightline (\autoref{fig:Sec3dLOS}). 
    Hence, studies that find large metallicity, temperature or density discrepancies between individual components \citep{Zahedy2019, Nielsen2022, Sameer2022} may be tracing gas arising from different origins or separated by large physical distances. 
    $\logNHIunit < 19.0$ \ion{H}{i} absorbers with velocity separations $|\Delta v_{\rm LOS}| < 20$ \kms\ can trace gas that is several Mpc away. 
    However, we find that 75 per cent of \ion{H}{i} absorbers with column densities $\logNHIunit > 16.0$ can be found within $\pm 150$ \kms\ of $M_* = 10^{10}$ M$_\odot$ central galaxies. 
    This value decreases to 70 and 120 \kms\ for $10^{8}$ and $10^{9}$ M$_\odot$ galaxies, respectively and increases to $250$ \kms\ for $10^{11}$ M$_\odot$ galaxies. 
    \item The fraction of absorbers that are associated with the central galaxy decreases with impact parameter (\autoref{fig:Sec4SATfcover}). 
    This decline is steepest for higher \ion{H}{i} column densities and lower stellar masses. 
    From impact parameters $b > 0.5R_{\rm vir}$, satellite and other halo galaxies begin to dominate the fraction of absorbers. 
    Contributions from the intergalactic medium are marginal, particularly for higher $M_*$ central galaxies and larger $N_{\ion{H}{i}}$. 
    However, we see in \autoref{fig:Sec4NHIb} that $> 80$ per cent of absorbers with column densities $13.0 < \logNHIunit < 16.0$ trace gas in the IGM. 
    \item Satellite galaxies with stellar mass $M_* < 10^{8}$ M$_\odot$ can contribute $\sim$40 per cent of the total sightlines that intersect satellites of a $10^{9}$ M$_\odot$ central galaxy. 
    This fraction reduces to $\sim$10 per cent for $10^{10}$ and $10^{11}$ M$_\odot$ centrals.
    For secondary haloes down the line of sight, 25 per cent of absorbers are attributed to $M_* < 10^{8}$ M$_\odot$ galaxies. 
    These findings are a possible explanation of \ion{H}{i} absorbers that do not have detected galaxy counterparts near the background source in observational studies. 
    \item After modelling the azimuthal angle of absorbers in a similar manner to observers, we find that the relative incidence of outflows compared to inflows increases as we move towards the minor axis. This signal is strongly dependent on the impact parameter and central galaxy stellar mass; at smaller $b$ and larger $M_*$, inflows begin to dominate at all azimuthal angles. 
    The larger incidence in simulations may be attributed to the singular classification of absorbers as either outflowing or inflowing; in TNG50, there is typically both inflowing and outflowing gas found along each sightline. 
    \item The median metallicity of absorbers increases towards the minor axis, consistent with the findings of previous theoretical studies \citep{Peroux2020, vandevoort2021} and observations \citep{Wendt2021}. When decomposing the signal into individual gas flows, we find that the increasing incidence of outflows does not drive the increasing metallicity, but rather both inflows and outflows increase in $Z$ towards larger azimuthal angles.  
    \item We find that analysing absorber-galaxy systems from the MUSE-ALMA haloes survey using the position of absorbers with respect to galaxies in $\Delta v$-$b$ space is useful to statistically associate the gas with its surrounding galaxies. Instead of assuming the galaxy nearest the absorber is the host galaxy of the gas, we suggest these $\Delta v$-$b$ diagrams (\autoref{fig:Sec6MAHadvise}) are a useful diagnostic that takes into account both the line of sight velocity difference and impact parameter. 
    \item In \autoref{fig:Sec6MAHFracCompare}, we show a direct comparison between the distribution of galaxies around absorbers in MUSE-ALMA Haloes and TNG50. The results are broadly consistent and highlight that observers may need to consider absorbers not bound to any galaxy as a plausible source of gas being probed, particularly around larger haloes. 
    \item There is a marginal ($\lesssim 0.5$ dex) difference in the peaks of the inflowing and outflowing gas metallicity distributions. The difference decreases for larger central galaxy stellar masses and increases for lower \ion{H}{i} column densities but never reaches the large differences seen in observational studies. In line with previous simulations \citep{Hafen2017, Rahmati2018}, we find that the observed bimodal metallicity distribution is not seen as strongly even after accounting for the various physical origins of the gas. 
\end{itemize}

We expect to expand this study to other common ions in absorption-line studies such as \ion{Mg}{ii}, \ion{C}{iv} and \ion{O}{vi} and to higher redshifts where more recent surveys are searching for galaxy counterparts to absorbers \citep[e.g.][]{Diaz2021, Bordoloi2023}. 
Upcoming surveys with instruments such as the Dark Energy Spectroscopic Instrument \citep[DESI; ][]{DESI1, DESI2}, WEAVE \citep{WEAVE} and 4MOST \citep{4MOST, 4MOSTcomm, Peroux2023} will produce millions of absorber-galaxy pairs and their results require a statistical interpretation. 
With the rapid proliferation of studies of galaxies surrounding absorbers of varying species and redshifts, it becomes more important than ever to interpret the results of these observational studies using simulations. 

\section*{Acknowledgements}
This research was supported by the International Space Science Institute (ISSI) in Bern, through ISSI International Team project \#564 (The Cosmic Baryon Cycle from Space). 
This research is supported by an Australian Government Research Training Program (RTP) Scholarship.
EMS and SW acknowledge the financial support of the Australian Research Council through grant CE170100013 (ASTRO3D).
DN and RR acknowledge funding from the Deutsche Forschungsgemeinschaft (DFG) through an Emmy Noether Research Group (grant number NE 2441/1-1). RR is a Fellow of the International Max Planck Research School for Astronomy and Cosmic Physics at the University of Heidelberg (IMPRS-HD). The TNG50 simulation was run with compute time awarded by the Gauss Centre for Supercomputing (GCS) under GCS Large-Scale Project GCS-DWAR on the Hazel Hen supercomputer at the High Performance Computing Center Stuttgart (HLRS). Additional computations were carried out on the Vera machine of the Max Planck Institute for Astronomy (MPIA) operated by the Max Planck Computational Data Facility (MPCDF).

\section*{Data Availability}

The IllustrisTNG simulations, including TNG50, are publicly available and accessible at \url{www.tng-project.org/data}, as described in \cite{Nelson2019a}. Data directly related to this publication is available on request from the corresponding author.

\bibliographystyle{mnras}
\bibliography{refs}

\appendix 
\section{Probability that the central galaxy hosts an absorber}
{Current \ion{H}{i} Ly$-\alpha$ surveys span a large variety of column densities and redshifts. 
In Sections \ref{sec:where} and \ref{sec:sat}, we highlight the challenges when attributing absorbers to a single galaxy. 
The commonly-adopted approach of assuming that the galaxy at lowest impact parameter hosts the absorber is far too simplistic. 
In \autoref{fig:Sec6MAHadvise}, we showed how the probability that an absorber belongs to a central galaxy varies with $b$ and $\Delta v$.  
We have only considered central galaxies with stellar masses $M_* = 10^9$, $10^{10}$ and $10^{11}$ M$_\odot$ and $\logNHIunit > 18.0$ absorbers at $z = 0.5$ which are most relevant for the MUSE-ALMA Haloes survey. 
Here, we extend the plots to more specific bins in \ion{H}{i} column density (pLLS, LLS, sub-DLA and DLA) and also include $M_* = 10^8$ M$_\odot$ central galaxies. 
We will expand the scope of this work to larger redshifts and \ion{Mg}{ii}, \ion{C}{iv} and \ion{O}{vi} absorbers in a forthcoming paper. 
}

\begin{figure*}
    \includegraphics[width=1.0\textwidth]{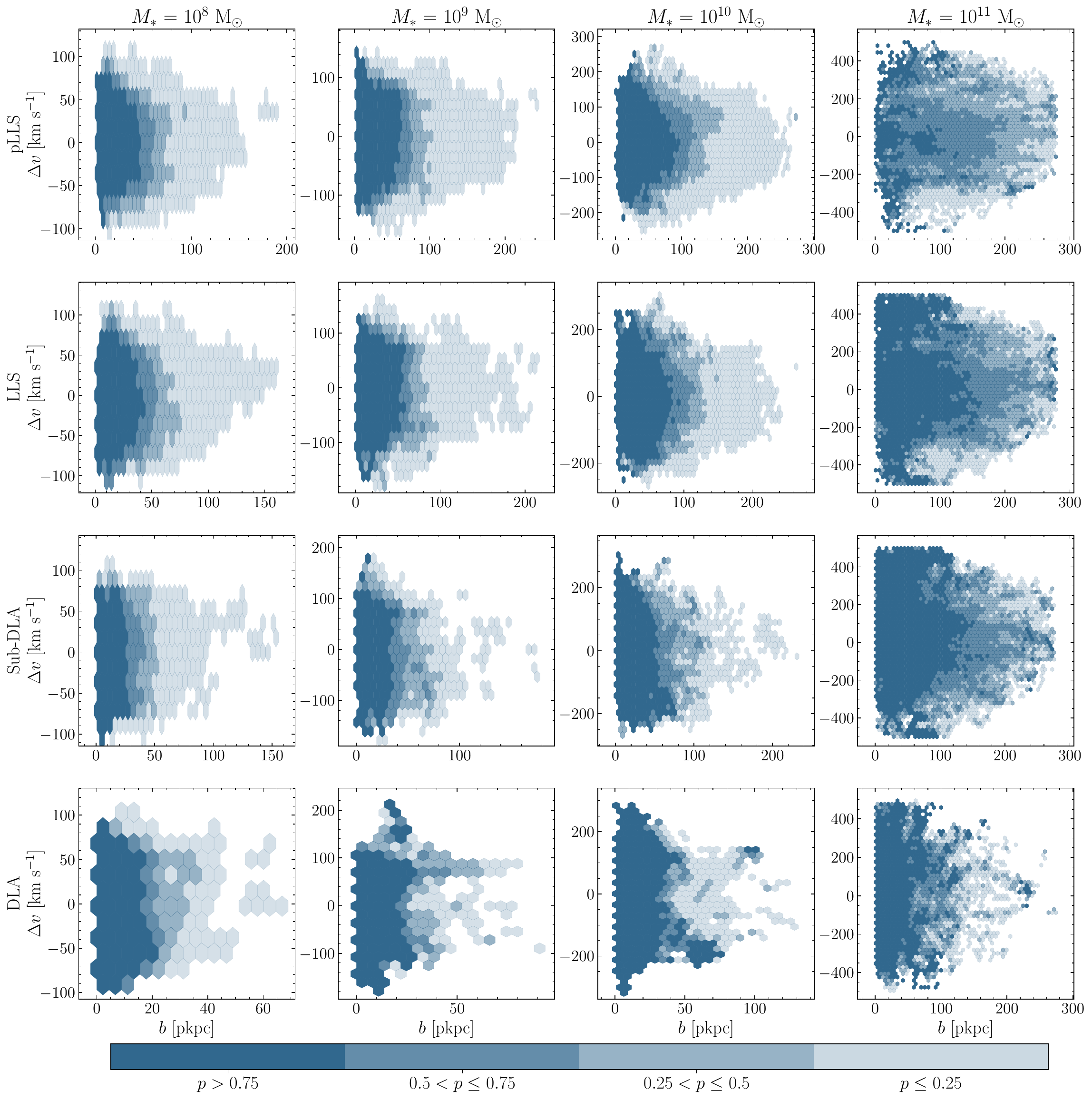}
    \caption{{An extension of the bottom row of \autoref{fig:Sec6MAHadvise} to include $10^8$ M$_\odot$ central galaxies and more specific column density bins. 
    Each hexbin is shaded by the probability, corresponding to the four ranges in the colourbar, that an absorber of given \ion{H}{i} column density (rows) is associated with a central galaxy of given stellar mass (columns). 
    We only include absorbers at $z = 0.5$, although we plan an extension to higher redshifts and different ions in a future work. }
    }
    \label{fig:AppProb}
\end{figure*}

\bsp	
\label{lastpage}
\end{document}